\documentclass[12pt,reqno]{article}
\usepackage{setspace,graphicx,amssymb,amsmath,latexsym,amsfonts,amscd,amsthm,multirow,ctable,mathdots,caption,array,diagbox,mathtools}
\usepackage{chet}
\usepackage{multicol}
\usepackage{tabularx,cite,mathrsfs}
\usepackage{authblk}
\usepackage{color}

\newcommand{\es}[2] {\begin{equation} \label{#1} \begin{split} #2 \end{split} \end{equation}}

\usepackage{fullpage}
\usepackage{stmaryrd}
\usepackage{rotating}

\usepackage{hyperref}

\usepackage{bbm}

\newcommand{\be}{\begin{eqnarray}}
\newcommand{\ee}{\end{eqnarray}}
\newcommand{\eeq}{\end{equation}}
\newcommand{\beq}{\begin{equation}}

\usepackage{simplewick}

\allowdisplaybreaks \numberwithin{equation}{section}
\DeclareSymbolFont{AMSa}{U}{msa}{m}{n}
\DeclareSymbolFont{AMSb}{U}{msb}{m}{n}
\DeclareMathSymbol{\fieldR}{\mathalpha}{AMSb}{"52}

\newcommand{\CO}{{\cal O}}

\newcommand{\tr}{{\rm tr}}

\def\beq{\begin{equation}}
\def\eeq{\end{equation}}
\def\bea{\begin{eqnarray}}
\def\eea{\end{eqnarray}}

\def\<{\langle}
\renewcommand\>{\rangle}
\newcommand\nn{\nonumber}

\def\X{\mathcal{X}}

\begin{document}

\setstretch{1.2}

\title{\vspace{-65pt}
\vspace{20pt}
    \textsc{\huge{
    Constraints on Flavored 2d CFT Partition Functions}
    \\
    }}

\author[a c]{Ethan Dyer}
\author[b]{A. Liam Fitzpatrick}
\author[b]{Yuan Xin}
\affil[a]{Stanford Institute for Theoretical Physics,
Via Pueblo, Stanford, CA 94305, USA}
\affil[b]{Boston University Physics Department,
Commonwealth Avenue, Boston, MA 02215, USA}
\affil[c]{Department of Physics and Astronomy, Johns Hopkins University,
Charles Street, Baltimore, MD 21218, USA}

\date{}

\abstract{
We study the implications of modular invariance on 2d CFT partition functions with abelian or non-abelian currents  when chemical potentials for the charges are turned on, i.e. when the partition functions are ``flavored''.  We begin with a new proof of the transformation law for the modular transformation of such partition functions.  Then we proceed to apply modular bootstrap techniques to constrain the spectrum of charged states in the theory.  We improve previous upper bounds on the state with the greatest ``mass-to-charge'' ratio in such theories, as well as upper bounds on the weight of the lightest charged state and the charge of the weakest charged state in the theory.  We apply the extremal functional method to theories that saturate such bounds, and in several cases we find the resulting prediction for the occupation numbers are precisely integers.  Because such theories sometimes do not saturate a bound on the full space of states but do saturate a bound in the neutral sector of states, we find that adding flavor allows the extremal functional method to solve for some partition functions that would not be accessible to it otherwise.
}

\maketitle

\clearpage

\tableofcontents

\newpage
\section{Introduction}

Modular invariance is a powerful tool for studying two-dimensional Conformal Field Theories (CFTs).  It is also a special case of crossing symmetry of CFT correlation functions \cite{Lunin:2000yv}, so aside from its intrinsic interest it is  useful as a simpler setting in which to explore many conformal bootstrap ideas and techniques \cite{Hellerman,Hellerman:2010qd,Friedan:2013cba}. 

A particularly appealing generalization of the conformal bootstrap equations is to consider correlation functions in the presence of nonlocal operators,  since this enlarges the set of CFT data that can be studied. In general, including nonlocal operators is a difficult problem, since their behavior under conformal transformations may be quite complicated.  However, one case where the problem remains tractable is when we consider modular invariance in the presence of a chemical potential in 2d CFTs.  A chemical potential corresponds to inserting the nonlocal operator  $y^{J_0} \equiv e^{2 \pi i z J_0}$ into the partition function,
\be
Z( \tau, z) \equiv \tr \left( q^{L_0- \frac{c}{24}} \bar{q}^{\bar{L}_0-\frac{c}{24}} y^{J_0} \right),
\label{eq:PF}
\ee
 where $J_0$ is the zero mode of a conserved current and $L_0, \bar{L}_0$ are Virasoro generators.  The resulting partition function is no longer modular invariant, but nevertheless has a well-defined and theory-independent transformation law  \cite{Kraus:2006wn}:
\be
Z(\frac{a \tau+b}{c \tau+d}, \frac{c z}{c \tau + d}) = e^{ \pi i k \left( \frac{ c z^2 }{c \tau+d} - \frac{ c \bar{z}^2}{c \bar{\tau}+d}\right)} Z(\tau, z) .
\label{eq:txn}
\ee  This transformation law
was used to constrain the spectrum of charges in general 2d CFTs in
\cite{Benjamin:2016fhe}. Proofs of (\ref{eq:txn}) so far
\cite{Kraus:2006wn,Dijkgraaf:1987vp,Dijkgraaf:1996xw,
  Benjamin:2016fhe} either are fairly complicated and technical or
else apply to special cases such as free boson constructions, and so
it is not clear what if any general lessons might be learned from
them.\footnote{We emphasize that we do not assume supersymmetry; additional techniques are available to proof the transformation law  for the elliptic genus  in the case of supersymmetric theories, see e.g.
\cite{krauel2012vertex}.} However, the very simple form of (\ref{eq:txn}) suggests it
should have an equally simple derivation.  Moreover, inserting the
nonlocal operator $y^{J_0}$ is equivalent to turning on a background
gauge field $A^\mu$ coupled to the conserved current $J^\mu$, which
suggests that one might be able to prove (\ref{eq:txn}) by studying
the CFT's effective action for $A_\mu$.  We will begin this paper in
section 2 by providing such a proof, and its generalization to a
non-abelian symmetry current $J^{a \mu}$.

Starting in section 3, we perform several analyses of the constraints
that follow from (\ref{eq:txn}) and its non-abelian generalization
using linear programming and semi-definite programming methods.  Our
main results are as follows.

\subsection*{Abelian Bounds}

We begin by reproducing, and improving the results of
\cite{Benjamin:2016fhe}, bounding properties of theories with an
abelian current. We place an upper bound on the dimension of the
lightest charged state, \es{bound}{ \Delta_{*} = \frac{c}{\alpha} +
  \mathcal{O}(1)\,, \ \ \ \alpha > 8\,.  } This bound is qualitatively similar to than
the bounds in \cite{Friedan:2013cba,Collier:2016cls} of non-charged
states.

We also improve the bound on the smallest ``mass-to-charge'' ratio in
the theory. These bounds are qualitatively related to the Weak Gravity
Conjecture (WGC), though gauge fields in the gravity duals are
Chern-Simons fields rather than Maxwell fields. Provocatively, we find
numerical evidence for a bound on the mass-to-charge ratio that scales
at large $c$ as $\sqrt{c}$,
consistent with the bulk gravity expectation.  This is stronger than
the bounds in \cite{Benjamin:2016fhe}, which scale as $c$. The
improvement again comes from increasing the number of derivatives of the characters used in the analysis.

Then, we discuss the bound on the charge gap $Q_*$. Without any further
assumptions, the numerical bound of charge gap is always $Q_*=1$ for
all $c$. We study two examples of $c=2$ and $8$ by turning on both a
gap in dimension $\Delta_*$ and in charge $Q_*$. There are kinks in
the $\Delta_*$ and $Q_*$ plots which can potentially be associated with full CFTs.

Lastly, we consider in detail spectra that extremize various gaps. We
use the extremal functional method, as well as extra information
contained in the charged spectra to study candidate theories at
$c=1$ and $8$. At $c=8$ we find that the level 1 $E_{8}$ Sugawara
theory saturates both the gap in dimension (as was found in
\cite{Collier:2016cls}) and in charge. Using this, we are able to
reproduce the full low lying spectrum, including charge assignments.

\subsection*{Non-abelian Bounds}

When the symmetry current $J^a$ is non-abelian, it is more appropriate to consider bounds on the dimensions of different representations in the theory.  We will mainly focus for specificity on the case where the gauge group $G$ is $SU(2)$ and the level $k$ is $1$, though our methods easily generalize to any algebra and level; the main advantage of $k=1, G=SU(2)$ is that convergence is fastest here, so our numerical results are most precise.

We first obtain bounds on the gap to all non-vacuum states in non-abelian theories. As the extended symmetry imposes additional relations on the spectrum, one may have hoped for stronger bounds. The results, however, are similar to those found in the abelian, or even non-flavored case. In particular, at large $c$ we find a bound of the form,
\es{nadimbound}{
\Delta_{*} = \frac{c}{\alpha} + \mathcal{O}(1)\,, \ \ \ \alpha > 8\,. 
} 

The real power of the modular constraints on the flavored partition function come from the ability to impose constraints independently on different representations. Taking advantage of this, we search for a bound on the gap to non-vacuum states transforming in the trivial representation, with no constraints imposed for other representations. At small $c$, there are interesting ``kinks'' at values of $c$ where the bounds on the gap in the neutral sector is minimized.  We focus on the case $c=3$, and use the extremal functional methods to find the low lying degeneracies at this kink. Reassuringly, we find integer degeneracies. This numerical spectral information allows us to guess an exact partition function saturating the bound in the neutral sector. In fact we find multiple partition functions are allowed if we are somewhat liberal in what spins are allowed for states in the theory.\footnote{The partition functions found are not strictly modular invariant, but invariant under a subgroup generated by $S$ and $T^{n}$ for $n=2$ or $n=4$.}  If all states must have integer or half-integer spins, we find a unique partition function,
\es{intpart}{
Z(\tau, \bar{\tau}, z, \bar{z})&= \frac{1}{4} \sum_{a,b,a',b'=0}^1 (-1)^{a b' + a' b} \left|\theta\left[^a_b\right](\tau,\frac{z}{2})\right|^4 | \theta[^{a'}_{b'}](\tau,0)|^2\,.
}
Allowing quarter-integer spins leads to multiple allowed partition functions that maximize the gap.  

Perhaps the most interesting aspect of this analysis however is not the specific partition function for this case, but rather that fact that searching for constraints in a representation dependent manner yields structure hidden to a flavor-blind analysis. 
This means that the extremal functional analysis allows one to ``discover'' a larger class of partition functions when flavored information is included than when it is not.  Moreover, uncovering flavored information can potentially split the degeneracy between theories with the same spectrum and therefore the same partition function, allowing us to address the age-old question of whether one can ``taste the shape of a drum.''  

Finally, we continue to refine our representation dependent analysis. For the case of $SU(2)_{1}$ we prove analytically that the theory either contains all representation, or the partition function splits into a product of the diagonal Sugawara partition function and a neutral, modular invariant partition function.

{\it After this work was completed, the paper \cite{Bae:2017kcl} appeared on arXiv also considering modular bootstrap constraints on theories with conserved currents, though the analysis there did not use the flavored partition function.}

\section{Partition Function Transformation and Background Gauge Fields}
\label{sec:partitionFunctionAndBkgdFields}

In this section, we will present an argument for the transformation law (\ref{eq:txn}) based on the effective action obtained upon integrating out the CFT in the presence of a background gauge field $A_\mu$.  Previous treatments  have pointed out that in the present context there are two different notions of the partition function that are natural.  One of these is the canonical partition function $Z(\tau,z)$ defined by (\ref{eq:PF}). Following \cite{Kraus:2006wn}, we will refer to an alternate definition as the ``path integral'' $Z_{\rm PI}(\tau, z)$:
\be
Z_{\rm PI}(\tau, z) \sim e^{ \pi k B(\tau, z) } Z(\tau, z) , \qquad
B(\tau, z) \equiv  \frac{z^2+\bar{z}^2}{2{\rm Im}(\tau)}.
\label{eq:PIvsPF}
\ee
Under $z' = \frac{z}{c \tau+d}, \tau' = \frac{a \tau+b}{c \tau+d}$, the factor $B$ is easily seen to transform as
\be
B(\tau',z') &=& B(\tau,z) - i  \left(  \frac{ c z^2}{c\tau+d}  - \frac{c \bar{z}^2}{c \bar{\tau}+d} \right).
\label{eq:extraPI}
\ee
The important point about the extra factor $B(\tau,z)$ is that its transformation cancels the transformation of $Z(\tau,z)$, leaving $Z_{\rm PI}$ invariant. The basic idea is that $Z_{\rm PI}$ should be the result of performing a path integral over the torus, and so should be modular invariant.  In free boson constructions, one can explicitly see how this factor is generated by the Legendre transform from the Lagrangian to the Hamiltonian \cite{Kraus:2006wn}.  
 However, we would like to see how this arises in a general CFT, without making any reference to a specific form of a Lagrangian.  We will begin with the case of an abelian current, and then consider the generalization to a non-abelian symmetry.

\subsection{Modular transformation and the ground state energy}

First, let us discuss in more detail how to define the ``path integral'' function $Z(\tau,z)$, what ambiguities are allowed in this definition, and why they do not affect the transformation law (\ref{eq:txn}).  In order to be invariant under  modular transformations, we will need to define the path integral to be invariant under diffeomorphisms and rigid rescalings $w \rightarrow \lambda^{-1} w$: 
\be
d \Psi e^{ - S_{\tau}[\Psi]} = d\Psi' e^{ - S_{\tau'} [\Psi']}.
\label{eq:meas}
\ee
Here, $\Psi$ are all the fields of the CFT.  As we review in appendix \ref{app:PItxn}, these two symmetries are sufficient to imply that the  path integral defined as an integral over this measure,
\be
Z_{\rm PI}(\tau, z) &\equiv& \int d \Psi e^{- S_{\tau}[\Psi] -\frac{i }{2\pi} \int_\tau A_{\bar{w}} J^{\bar{w}}}, \qquad A_{\bar{w}} = -i \frac{z}{2 {\rm Im}(\tau)},
\ee
is invariant under modular transformations:
\be
Z_{\rm PI}\left( \frac{a \tau+b}{c \tau+d}, \frac{c z}{c\tau+d}\right) &=& Z_{\rm PI}(\tau, z).
\ee
Different choices of regulators will change $\log Z_{\rm PI}$ by local terms.  However, the local terms allowed by diffeomorphism invariance and scale invariance  do not affect the transformation law.  For instance, one can shift the effective action by a local term proportional to
\be
\int_\tau d^2 x \sqrt{g} A_\mu A^\mu \sim \int_\tau dw d\bar{w} A_w A_{\bar{w}}  \sim \frac{z  \bar{z}}{4 {\rm Im}(\tau)}.
\ee
This term arises in the difference between a regulator that preserves the vector current $J^\mu$ symmetry and one that preserves the axial current $\epsilon^{\mu\nu}J_\nu$.  However, it is easily seen to be both Weyl invariant and diffeomorphism invariant, and is invariant under modular transformations. So its coefficient is irrelevant for our purposes, and from now on we will neglect such terms without loss of generality.

Now, the next question is how do we relate the ``path integral'' $Z_{\rm PI}$ to the partition function $Z$? The key point is that turning on a background field $A_\mu$ not only turns on a chemical potential, but it can also shift the ground state energy, since at fixed $\beta$ such a shift affects only the overall normalization of the path integral.  

 In the example of the free boson, this energy shift is seen explicitly by doing a Legendre transform, but we can see it in full generality by considering the effective action for $A_\mu$.  To see the shift, it is sufficient to calculate the ground state energy, so we can take the limit of the torus where $\tau=  i \frac{\beta}{2 \pi} , \beta \gg 1 $.  In this limit, the torus becomes a cylinder, and the effective action is conformally related to that in flat space, where it is universal and known in closed form.  Including the action for a background metric as well, we can write
 \be
 \log Z =  \int  d^2 x \sqrt{-g} \left( \frac{c}{48 \pi } R \Box^{-1} R + \frac{k}{8 \pi} F^{\mu\nu} \Box^{-1} F_{\mu\nu} \right).
 \label{eq:logZcyl}
 \ee
Because of the inverse Laplacians, the mapping to the cylinder is a bit subtle.  For the metric contribution, it is easiest to work with the Wess-Zumino anomaly action directly, $S_{\rm WZ} = \frac{c}{24\pi} \int d^2 x\sqrt{-g} \left(  \sigma R + (\partial \sigma)^2 \right)$, and take $\sigma(w) = w+\bar{w}$, which reproduces the standard ground state energy shift $-\frac{c}{12}$ from the Schwarzian derivative. By contrast, the gauge field term in (\ref{eq:logZcyl}) is invariant under Weyl transformations, and its contribution to the ground state energy just comes from evaluating the non-local term on the cylinder. To avoid ambiguities associated with the inverse Laplacian, it is clearest to use the fact that the effective action is the generating functional for the $J^\mu$ correlators, so we know that we can equivalently write the gauge field part of $\log Z$ as
\be
W_A[A^\mu] = \int \frac{d^2 x d^2 x'}{(2\pi i )^2} A_\mu(x)  A_\nu(x')  \< J^\mu(x) J^\nu(x') \>.
\ee
On the plane, $\< J^{\bar{w}}(w) J^{\bar{w}}(w')\> = \frac{k}{(w-w')^2}$.  Mapping to the cylinder and taking $A_{\bar{w}}$ to be constant, we have\footnote{We performed this integration as follows. 
First,  shift $w \rightarrow w+w'$ to eliminate $w'$ and immediately do the $d^2 w'$ integral, producing just a factor of the volume $2 \pi \beta$ of the torus.
Passing to $t,\theta$ coordinates:
\be
W_A[A_\mu] = - \frac{\beta k  A_{\bar{w}}^2}{2\pi} \int_{-\beta/2}^{\beta/2} dt  \int_0^{2\pi} d\theta\frac{1}{(e^{\frac{t+i \theta}{2}} -e^{\frac{-t-i \theta}{2}})^2} .
\ee
If we do the $\theta$ integral first, this vanishes, except when $t=0$ where it is divergent; the integral over $\theta$ is proportional to $\delta(t)$.  We avoid this subtlety if we do the $t$ integral first, in which case we obtain
\be
W_A[A_\mu] =- \frac{\beta k A_{\bar{w}}^2}{2\pi}  \int_0^{2 \pi} d \theta \frac{\sinh \frac{\beta}{2}}{\cos \theta - \cosh \frac{\beta}{2}} =   \beta k A_{\bar{w}}^2 .
\ee
}
\be
W_A[A_\mu] \rightarrow \frac{k}{(2\pi i)^2} \int d^2 w d^2 w' \frac{A_{\bar{w}}^2}{\left( e^{\frac{w-w'}{2}} - e^{\frac{w'-w}{2}}\right)^2} = \beta k A_{\bar{w}}^2.
\ee
Combining the above with a symmetric combination from $A_w$, we put everything together to obtain the ground state energy:
\be
E_0 = -\lim_{\beta \rightarrow \infty} \beta^{-1} \log Z_{\rm PI} = - \frac{c}{12} + \delta E , \qquad
\delta E = -k (A_w^2+ A_{\bar{w}}^2).
\ee
Therefore, the path integral differs from the canonical partition function by an extra factor $e^{-\beta \phantom{.} \delta E}$, which in turn produces the factor $- \pi k B(\tau,z)$ in (\ref{eq:PIvsPF},\ref{eq:extraPI}). 
So at last we see that this  factor is universally the contribution to the partition function from the shift in the ground state energy due to the background gauge field.

Summarizing, the canonical partition function $Z(\tau,z)$ in (\ref{eq:PF}) is defined to have a ground state energy $-\frac{c}{12}$.  However, any path integral over the torus using a regulator that preserves diffeomorphisms and rescalings will have a ground state energy equal to $-\frac{c}{12} - k (A_w^2 + A_{\bar{w}}^2)$, plus possible terms that do not affect the modular transformation of $Z(\tau,z)$. 

\subsection{Non-abelian current transformation}

The generalization to the case of a non-abelian is straightforward, and can be made as follows.  Unlike in the abelian case, the effective action is not quadratic.  However, we can write it formally as the sum over all connected diagrams:
\be
W_A[A^{a \mu} ] = \sum_{n=1}^\infty \left(\prod_{i=1}^n  \int \frac{d^2 w_i}{2\pi i} A^{a_i}_w \right) \< J^{a_1}(w_1) \dots J^{a_n}(w_n) \>_{\rm conn} .
\ee
As before, we want to set $A_w^a, A_{\bar{w}}^a$ to be constant on the cylinder and integrate over $d^2 w_i$.  For the part quadratic in $A$,
 the computation proceeds just as in the abelian case, we simply have an extra index for the different components of $J^a_w$. The background field couples as
 \be
 \frac{-i}{2\pi} \int_\tau A_\mu^a J^{\mu a}, \qquad A_w^a = -i \frac{\bar{z}^a}{2 {\rm Im}(\tau)}.
 \ee
$A$'s contribution to the ground state energy is
 \be
 \delta E \cong -k((A_w^a)^2+(A_{\bar{w}}^a)^2),
 \ee
 which transforms under modular transformations as
 \be
 -\beta \delta E \rightarrow -\beta \delta E - i \pi k  \left(  \frac{c (z^a)^2}{c\tau+d} - \frac{c (\bar{z}^a)^2}{c \bar{\tau}+d} \right).
 \ee

 That leaves the contribution from the  higher-point functions. We can always write these in terms of lower-point function by using the recursive formula
  \be
  J^a(w)J^b(0) \sim \frac{k \delta^{ab}}{w^2} + \frac{f^{abc} J^c(0)}{w},
  \ee
  where $\sim$ means `up to non-singular terms'.  The $\frac{k \delta^{ab}}{w^2}$ piece manifestly generates disconnected diagrams - it produces the two-point function times the $(n-2)$-point function - so it does not contribute to the effective action for higher-point correlators.  But, since we multiply the correlator by $A^a_w$ in the effective action, the $f^{abc}$ term also gives no contribution for constant $A_w^a$:
  \be
  A_w^a J^a(w) A_w^b J^b(0) \sim \frac{k A_w^2 }{ w^2} + A_w^a A_w^b f^{abc} \frac{J^c(0)}{w} = \frac{k A_w^2 }{w^2}
  \ee
  since $A^a A^b f^{abc} =0$.   Therefore only the two-point functions contribute.

\section{Modular Bootstrap with Chemical Potentials}
\label{sec:constraints}

\subsection{Basic setup}

In all the cases we consider, we will assume the presence of a
conserved current $J^a$ in the theory. In general, it is convenient to
separate the stress tensor $T$ of the theory into a Sugawara stress
tensor piece and a residual piece:
\be T^{(0)} \equiv T - T^{\rm sug},
\qquad T^{\rm sug} = \frac{1/2}{k+\widetilde{h}_G} \sum_{a=1}^{|G|} : J^a J^a : ,
\ee
where $\widetilde{h}_G$ is the dual Coxter number,
because the modes of $T^{(0)}$ commute with the modes of
$J^a$. Furthermore among themselves they form a Virasoro algebra with
central charge
\be c^{(0)} = c- c^{\rm sug}, \qquad c^{\rm sug} =
\frac{k |G|}{k+\widetilde{h}_G} .
\ee
Similarly, we can separate the
Virasoro generators $L_n = L_n^{(0)}+L_n^{\rm sug}$ and the weights
$h = h^{(0)} + h^{\rm sug}$ into a part that comes from $T^{(0)}$ and
a part that comes from $T^{\rm sug}$. For most representations, the
distinction between $T$ and $T^{(0)}$ will not make much difference,
since the partition function just counts states at each level.
However, for the special cases with shortening conditions, some
descendants becomes null and do not contribute to the partition
function, and this is easier to see using the modes of $T^{(0)}$.

The characters of the Kac-moody algebra 
\begin{align}\label{eq:kac-moodyAlgebra}
  \X_{\mathbf{\mu},k}(\tau,\mathbf{z})={\rm
  Tr}_{V_{\mathbf{\mu},k}}q^{L_0^{\rm sug}-\frac{c^{\rm sug}}{24}}
  e^{2\pi i \mathbf{z}\cdot \mathbf{H}_0}
\end{align}
are constructed by acting modes of $J^a$ 
on some highest weight state which has weight $\mathbf \mu$.\footnote{See e.g. \cite{DiFrancesco:1997nk}  for a standard introduction.}  Here, ${\bf H}_0$ is the vector of Cartan generators of the algebra.  In the case of an abelian symmetry, ${\bf H_0} = J_0$ and the characters for a generic primary are simply $ q^{ - \frac{c^{\rm sug}-1}{24}}e^{2 \pi i z Q}/\eta(\tau)$.  In the case of a non-abelian symmetry, the characters are more complicated.  Some descendants of such a highest weight state may be null so it is non-trivial to write down its form. However, for the purpose of the modular bootstrap, the only property of such characters we use is that the characters transform covariantly 
\begin{align}\label{eq:kac-MoodyCharacterModularTransformation}
  \X_{\mathbf{\mu},k}\left(-\frac{1}{\tau},\frac{\mathbf{z}}{\tau}\right)=
  e^{\frac{i \pi k \mathbf{z}^2}{\tau}}
  \sum_{\mathbf{\mu^\prime}}S_{\mathbf{\mu\mu^\prime}}^k\X_{\mathbf{\mu^\prime},k}(\tau,\mathbf{z}) ,
\end{align}
where the matrix $S$ depends on the symmetry group and level $k$.
These characters do not include the modes of $T^{(0)}$ yet. Since the algebra generated by modes of  $J^a$ is completely orthogonal to that generated by modes of $T^{(0)}$,
the character generated by the full extended algebra simply factorizes into a Kac-Moody character 
and a Virasoro character
\begin{equation}\label{eq:characterFactoraization}
  \X_{\mathbf{\mu},k,h}(\tau, \mathbf{z}) = \X_{\mathbf{\mu},k}(\tau,\mathbf{z}) \X_{h^{(0)}}(\tau)~.
\end{equation}
Like the simple Virasoro character, the character is different if the primary saturates the unitarity bound:
\begin{equation}\label{eq:virasoroPart}
  \X_{h^{(0)}} (\tau) = \left\{ \begin{array}{cc} \frac{q^{h^{(0)}-\frac{c^{(0)}-1}{24}}}{\eta(\tau)} & h^{(0)} >0 \\
\frac{(1-q)q^{-\frac{c^{(0)}-1}{24}}}{\eta(\tau)} & h^{(0)}=0 \end{array} \right. .
\end{equation}
 The same goes for the anti-holomorphic part.
The full partition function is
\begin{equation}\label{eq:partitionFunctionNonAbelian}
  Z(\tau, \bar{\tau}, \mathbf{z},\mathbf{\bar z}) =
  \sum_{\mathbf{\mu},\mathbf{\bar{\mu}},h,\bar{h}}
  d_{\mathbf{\mu},\mathbf{\bar{\mu}},h,\bar{h}}  \X_{\mathbf{\mu},k}(\tau,\mathbf{z})\X_{\mathbf{\bar \mu},k}(\bar{\tau},\bar{\mathbf{z}}) \X_{h^{(0)}}(\tau) \X_{\bar h^{(0)}}(\bar{\tau}) ~.
\end{equation}
In the above equation the $\bar{\mu}$ means the representation of the anti-holomorphic part.

When we are dealing with a non-abelian symmetry, it will be convenient to define a matrix $M_{\mu, \bar{\mu}}$ whose components are the coefficients of the contributions to the partition function from the different representations:
\es{eq:partitionFunctionAsAMatrix}{
  M(\tau,\bar{\tau})_{\mathbf{\mu},\mathbf{\bar{\mu}}}&= \sum_{h, \bar{h}}
   d_{\mathbf{\mu},\mathbf{\bar{\mu}},h,\bar{h}} \X_{h^{(0)}}(\tau) \X_{\bar h^{(0)}}(\bar{\tau})\\
     Z(\tau, \bar{\tau}, \mathbf{z},\mathbf{\bar z})&= \sum_{\mathbf{\mu},\mathbf{\bar{\mu}}} M(\tau,\bar{\tau})_{\mathbf{\mu},\mathbf{\bar{\mu}}}\X_{\mathbf{\mu},k}(\tau,\mathbf{z})\X_{\mathbf{\bar \mu},k}(\bar{\tau},\bar{\mathbf{z}})\,.
}
 Modular transformations on the partition function translates into a specific modular transformation of the matrix $M_{\mu, \bar{\mu}}$.  To see this transformation law, simply separate out the transformation law of $Z$ into its irrep constituents: 
\es{eq:modularTransformationOfPartitionFunction}{
  0&=Z(-\frac{1}{\tau}, -\frac{1}{\bar{\tau}}, \frac{\mathbf{z}}{\tau},\frac{\mathbf{\bar z}}{\bar{\tau}})  -
   e^{i \pi k \left(\frac{\mathbf{z}^2}{\tau}-\frac{\bar{\mathbf{z}}^2}{\bar{\tau}}\right)} Z(\tau, \bar{\tau}, \mathbf{z},\mathbf{\bar z}) \\
   &=\sum_{\mathbf{\mu},\mathbf{\bar{\mu}}} M\left(-\frac{1}{\tau},-\frac{1}{\bar{\tau}}\right)_{\mathbf{\mu},\mathbf{\bar{\mu}}}\X_{\mathbf{\mu},k}\left(-\frac{1}{\tau},\frac{\mathbf z}{\tau}\right)\bar{\X}_{\mathbf{\bar \mu},k}\left(-\frac{1}{\bar{\tau}},\frac{\bar{\mathbf{z}}}{\bar{\tau}}\right)-e^{i \pi k \left(\frac{\mathbf{z}^2}{\tau}-\frac{\bar{\mathbf{z}}^2}{\bar{\tau}}\right)}
   M(\tau,\bar{\tau})_{\mathbf{\mu},\mathbf{\bar{\mu}}}\X_{\mathbf{\mu},k}(\tau,\mathbf{z})\bar{\X}_{\mathbf{\bar \mu},k}(\bar{\tau},\bar{\mathbf{z}})\\
&=\sum_{\mathbf{\mu},\mathbf{\bar{\mu}}}\left(\sum_{\mathbf{\mu}^{\prime},\mathbf{\bar{\mu}^{\prime}}}S_{\mathbf{\mu^\prime\mu}}^kM\left(-\frac{1}{\tau},-\frac{1}{\bar{\tau}}\right)_{\mathbf{\mu^{\prime}},\mathbf{\bar{\mu}^{\prime}}}\bar{S}_{\mathbf{\bar{\mu}^\prime\bar{\mu}}}^k- M(\tau,\bar{\tau})_{\mathbf{\mu},\mathbf{\bar{\mu}}}\right)e^{i \pi k \left(\frac{\mathbf{z}^2}{\tau}-\frac{\bar{\mathbf{z}}^2}{\bar{\tau}}\right)}\X_{\mathbf{\bar \mu},k}(\bar{\tau},\bar{\mathbf{z}})\bar{\X}_{\mathbf{\bar \mu},k}(\bar{\tau},\bar{\mathbf{z}})\,,
}
where we have used the transformation rule, \eqref{eq:kac-MoodyCharacterModularTransformation}, and the definition (\ref{eq:partitionFunctionAsAMatrix}). Stripping off the characters, the above crossing equation is equivalent to a crossing equation for the matrix
\begin{equation}\label{eq:modularTransformationOfMatrix}
  0 = M(\tau,\bar{\tau})_{\mu,\bar \mu} - S^T_{\mu,\mu^\prime}
  M\left(-\frac{1}{\tau},-\frac{1}{\bar{\tau}}\right)_{\mu^\prime,\bar{\mu}^\prime}
  \bar{S}_{\bar{\mu}^\prime,\bar{\mu}}\,.
\end{equation}

For the constraints on theories with non-abelian currents, equation (\ref{eq:modularTransformationOfMatrix}) is the form of the constraint that we will use. For each bootstrap question we will input
the symmetry group and level $k$.

\subsection{Semidefinite Projective Functionals and the Extremal Method}

To be self-contained, we will briefly review linear and semidefinite programming methods as applied to the modular bootstrap; for more thorough reviews and some examples of applications, see e.g. \cite{Friedan:2013cba,Collier:2016cls,Qualls:2013eha,Qualls:2015bta,Collier:2017shs,Montero:2016tif,Heidenreich:2016aqi,Das:2017vej}, or \cite{ElShowk:2012hu,Rychkov:2016iqz,Simmons-Duffin:2016gjk,Rattazzi:2008pe,Kos:2014bka,Poland:2011ey,Esterlis:2016psv} for reviews and some of the original papers developing methods in the standard bootstrap that we will adopt directly.  The starting point is equation (\ref{eq:modularTransformationOfPartitionFunction}), which can be written
\be
0 &=& \sum_{h, \bar{h}, \mu', \bar{\mu}' } d_{\mu, \bar{\mu}, h, \bar{h} } \left( F_{\mu, \bar{\mu}, h, \bar{h}}\right)_{\mu', \bar{\mu}'}(\tau, \bar{\tau}) , 
\label{eq:ModBootEq2} \\
 \left( F_{\mu, \bar{\mu}, h, \bar{h}}\right)_{\mu', \bar{\mu}'}(\tau, \bar{\tau})  &\equiv&  \left[ \delta_{\mu \mu'} \delta_{\bar{\mu} \bar{\mu}'} \X_{h^{(0)}}(\tau) \X_{\bar{h}^{(0)}}(\tau) - S_{\mu' \mu} S_{\bar \mu', \bar{\mu}} \X_{h^{(0)}}(-1/\tau) \X_{\bar{h}^{(0)}}(-1/\bar{\tau}) \right] .\nn\\
\ee
The occupation numbers $d_{\mu, \bar{\mu}, h,\bar{h}}$ are all non-negative, and include in particular the vacuum $d_{\rm vac}=1$. One is generally interested in proving that there exist states in the theory with various properties, for instance that there exists a state with $\Delta < \Delta_{\rm max}$ for some value of $\Delta_{\rm max}$.  Let us abstractly call a choice of such properties ``$P$''.  Then, one can prove that there is at least one state in the theory with properties $P$ as long as one can find a linear functional $\rho$ that maps the characters to real numbers such that it is positive on the vacuum and also positive on all states not satisfying $P$.  In equations,
\be
\rho({\rm vac})= 1 , \textrm{ and }  \rho(F_{\mu, \bar{\mu}, h, \bar{h}} ) \ge 0 \textrm{ unless } (\mu, \bar{\mu}, h, \bar{h}) \textrm{ satisfies } P.
\ee
 The normalization $\rho({\rm vac})=1$ is conventional.  If such a linear functional $\rho$ exists, then there must be a state in the theory with the properties $P$, otherwise $\rho$ acting on equation (\ref{eq:ModBootEq2}) would imply $0 \ge 1$.

Usually we will be interested in not just one choice of $P$ but a continuous family of choices  $P_s$ parameterized by a continuous variable (or variables) $s$. Typically, $s$ will be something like the bound $\Delta_{\rm max}$ in the example above, so that as one decreases $s$ the set of states with  property $P_s$  grows and therefore the set of linear functionals that are positive on all such states shrinks.  Critical values $s_*$ of $s$ where the set of such linear functionals vanishes are especially interesting: aside from giving the best possible bounds, at these points one can use the ``extremal functional method'' \cite{ElShowk:2012hu} to determine all of the occupation numbers $d_{\mu, \bar{\mu}, h, \bar{h}}$.  The basic idea behind this is that for any $s$, the space of functions $F_{\mu, \bar{\mu}, h, \bar{h}}$ spanned by states satisfying $P_s$ is a polytope where $-F_{\rm vac}$ is inside the polytope for $s<s_*$ and outside the polytope for $s>s_*$.  At exactly $s_*$, $-F_{\rm vac}$ passes through one of the faces of the polytope, so there is a unique positive semidefinite linear combination of the states satisfying $P_{s_*}$ that cancels the contribution from the vacuum in (\ref{eq:ModBootEq2}).  In practice, we have to work with finite-dimensional projections of the full space of functions $F_{\mu, \bar{\mu}, h, \bar{h}}$, but one optimistically expects to converge to a unique solution as the dimensionality of the projected space increases.  We will encounter some exceptions that we will discuss as we come to them.

\section{Abelian Bounds}

In this section, we perform a systematic numerical analysis on the bounds on the gap in dimensions and charges, as well as on the smallest charge-to-mass ratio allowed in a theory with a $U(1)$ current.  

\subsection{Semi-definite Programming with Continuous Charge $Q$}

For the abelian case, for simplicity we will not use the full Kac-Moody characters, but rather just the Virasoro characters $\chi_h(q)$:
\begin{align}
  Z(\tau, z) = 
  \sum_{h, \bar h, Q, \bar Q} d_{Q, \bar{Q}, h , \bar{h}} y^Q \bar{y}^{\bar{Q}}  \chi_h(q) \chi_{\bar{h}}(\bar{q}).
\end{align}
where $q= e^{2\pi i \tau}$, $\bar q = e^{-2\pi i \bar \tau}$, $y =
e^{2\pi i z}$, $\bar y = e^{-2\pi i \bar z}$.  

We will consider left-right symmetric theories with $c = \bar c$, and for simplicity we set
$\bar z = 0$.

As in \cite{Friedan:2013cba}, we reduce the characters using
the $S$ invariant factor
$|\tau|^{\frac{1}{2}}|\eta(\tau)|^2$. Furthermore we reduce the
partition function as
\begin{align}
  \hat Z(\tau, z) \equiv  |e^{\frac{i\pi z^2}{2\tau} }|^2
  |\tau|^{\frac{1}{2}}|\eta(\tau)|^2 Z(\tau, z) ,
\end{align}
so that $\hat Z(\tau, z)$ is invariant under $ \left( \tau \mapsto
  -\frac{1}{\tau}, z \mapsto \frac{z}{\tau} \right)$. 
The characters are reduced into
\begin{align}
  \hat \chi_0(q) = e^{\frac{i\pi z^2}{2\tau} } \tau^{\frac{1}{4}}
  q^{-\frac{c-1}{24}}(1-q) ,~~~~
  \hat \chi_h(q) y^Q = e^{\frac{i\pi z^2}{2\tau} } \tau^{\frac{1}{4}}
  q^{-\frac{c-1}{24}}q^h y^Q .
\end{align}
We consider linear functionals of the form
\begin{align}
  \rho \equiv \sum_{m+n+k/2 {\rm ~odd,~} k {\rm ~even}}
  \left. \alpha_{m,n,k} \ \partial_{t}^m \, \partial_{\bar t}^n\, 
  \partial_{w}^k
  \right|_{t=\bar t=w=0},
\end{align}
where the change of variable  $\tau = i e^{t}$ and $z =
we^{\frac{t}{2}}$ is made so that $t \mapsto -t$ and $w^2 \mapsto
-w^2$ under $S$ transformations.\footnote{In this expression for $\rho$, we have not used any $\bar{z}$ derivatives and so do not use information about the anti-holomorphic charge $\bar{Q}$.  This is mainly for simplicity and efficiency; it is in principle straightforward, though more computationally intensive, to use $\bar{Q}$ information as well.  Later in this section we will in fact perform one analysis where we keep $\bar{z}$ derivatives to demonstrate this point.}

We arrive at a two variable functional
\begin{align}
  \rho_l(\Delta,Q) &\equiv \rho\left[ \hat\chi_{\frac{\Delta-l}{2}}(\tau)
  \hat{\bar\chi}_{\frac{\Delta+l}{2}}(\bar{\tau}) y^Q +
  \hat{\chi}_{\frac{\Delta+l}{2}}(\tau)
  \hat{\bar\chi}_{\frac{\Delta-l}{2}}(\bar \tau)  \right] , \nn \\
  \rho({\rm vac}) &\equiv \rho\left[ \hat\chi_{0}(\tau)
  \hat{\bar\chi}_{0}(\bar{\tau}) \right],
\end{align}
where we assume the spectrum is parity symmetric.

Part of the challenge of the abelian analysis is that we do not assume
charge quantization, i.e. technically we allow the gauge group to be
$\mathbb{R}$ instead of $U(1)$, which means that we have to deal with
not just one but two continuous parameters, $\Delta$ and $Q$.  This
complicates the application of positive semi-definite approaches,
since these are based on constructing positive functionals of the
characters and in general the space of such functionals is more
complicated for multiple variables than for a single variable.  In
particular, for a single variable, positive semi-definite functionals
can be written without loss of generality as a sum of squares plus
a linear term times a sum of squares.  For multiple variables, such a parameterization is no
longer completely general.  One way to deal with this issue is simply
to discretize in, say, $Q$, but we find that such an approach becomes
difficult to implement in practice since the discretization needs to
become very fine to prevent the numeric search from picking
functionals that become negative in between the discretization points.
The approach we take is instead to limit the search space to functionals
that are still a sum of squares plus a linear term times sums of squares. In the limit of very
high order polynomials, one might expect that such functionals can approximate the extremal functionals arbitrarily well.  In any case, while such
functionals might not give the {\it best} possible bounds, they
nevertheless produce valid bounds.  

Even restricting to polynomial functionals, there remains a practical problem of how to implement the search over such functionals using available software for semi-definite programming analyses.  In appendix \ref{sec:multivariate}, we discuss how to massage this problem into an appropriate form for use with SDPB.

\subsection{Bound on Dimension of Lightest Charged State}
\label{sec:boundDimension}
\begin{figure}[ht]
  \centering
  \includegraphics[width=0.6\textwidth]{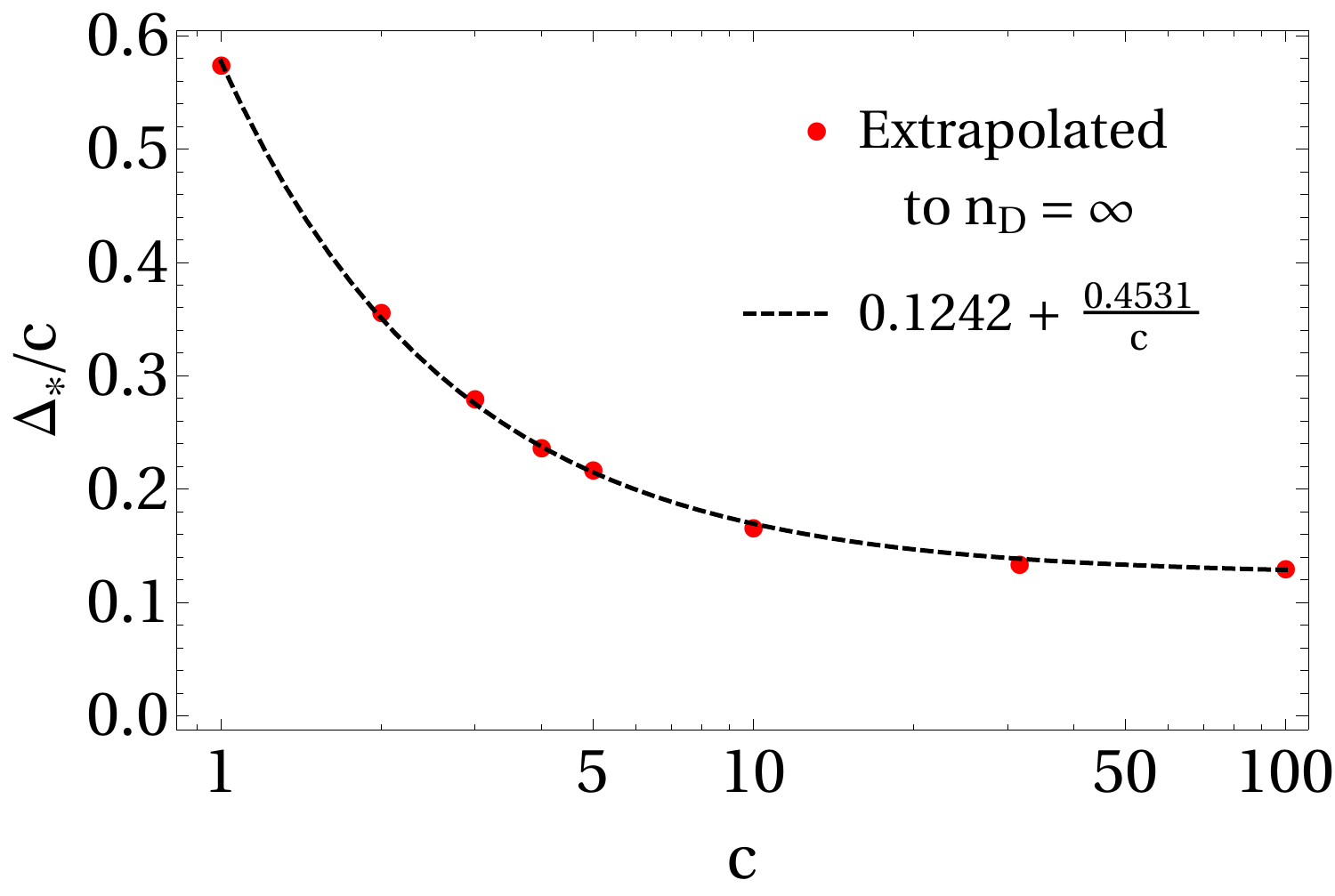}
  \caption{\label{fig:dimgapExtrapolate} Bounds of the dimension of
    lightest charged state assuming the theory has U(1) symmetry.  The
    extrapolated gaps at $n_D \rightarrow \infty$ with the trend
    line.}
\end{figure}

With the flavored partition function we can bound the dimension
$\Delta_*$ of lightest charged state in any theory with a U(1).  The
bound for different $c$ is summerized in Table.~\ref{tab:dimgap}. We
extrapolate the bound values to $n_D \rightarrow \infty$ using a
linear function of $\frac{1}{n_D}$ similar to what is done in
\cite{Collier:2016cls} for $c \leq 100$ where the convergence of the
bound values is significant. Then we extrapolate the bounds to
$n_D\rightarrow \infty$ and $c \rightarrow \infty$ by fitting the finite $n_D$  and $c$ results to  a linear
function of $\frac{1}{c}$ and $1/n_D$ and extrapolating, as shown in
Fig.~\ref{fig:dimgapExtrapolate}.  In this test we take
$\tau = \frac{i \beta}{2 \pi}$ with $\beta$ real to avoid the
complication of spin. It is not understood a priori why the
form of this fit works, but empirically it agrees well with the data at large $c$ and $n_D$.

\begin{table}[ht]
  \begin{center}
    $\begin{array}{cccccc} \hline
       n_D & c=1 & c=2 & c=3 & c=4 & c=5 \\
       \hline
       5 & 0.60000 & 0.77500 & 0.95000 & 1.2000 & 1.3000 \\
       7 & 0.58750 & 0.76250 & 0.93750 & 1.1000 & 1.2750 \\
       9 & 0.58047 & 0.73438 & 0.88750 & 1.0375 & 1.2000 \\
       11 & 0.58037 & 0.73281 & 0.88125 & 1.0312 & 1.1750 \\
       13 & 0.57988 & 0.72969 & 0.87187 & 1.0188 & 1.1625 \\
       15 & 0.57983 & 0.72812 & 0.87031 & 1.0125 & 1.1500 \\
       17 & 0.57958 & 0.72734 & 0.86875 & 1.0062 & 1.1437 \\
       19 & 0.57958 & 0.72656 & 0.86719 & 1.0047 & 1.1406 \\
       21 & 0.57957 & 0.72637 & 0.86641 & 1.0031 & 1.1375 \\
       23 & 0.57956 & 0.72627 & 0.86602 & 1.0023 & 1.1367 \\
       25 & 0.57956 & 0.72617 & 0.86563 & 1.0016 & 1.1352 \\
       27 & 0.57956 & 0.72615 & 0.86553 & 1.0012 & 1.1344 \\
       29 & 0.57956 & 0.72610 & 0.86533 & 1.0008 & 1.1336 \\
       \hline
       \hline
       n_D & c=10^1 & c=10^{\frac{3}{2}} & c=10^2 & c=10^{\frac{5}{2}} &
                                                                         c=10^3 \\
       \hline
       5 & 2.40 & 6.40 & 19.2 & 102. & 179.2 \\
       9 & 2.00 & 5.60 & 16.0 & 51.2 & 166.4 \\
       13 & 1.90 & 5.00 & 15.6 & 49.6 & 160.0 \\
       17 & 1.85 & 4.90 & 15.2 & 48.8 & 158.4 \\
       21 & 1.80 & 4.80 & 14.8 & 47.2 & 155.2 \\
       25 & 1.79 & 4.70 & 14.4 & 46.4 & 153.6 \\
       29 & 1.79 & 4.60 & 14.0 & 46.0 & 152.0 \\
       33 & 1.78 & 4.55 & 13.8 & 45.2 & 150.4 \\
       37 & 1.78 & 4.53 & 13.6 & 44.8 & 148.8 \\
       41 & 1.77 & 4.50 & 13.5 & 44.4 & 148.0 \\
       \hline
     \end{array}
     $
   \end{center}
   \caption{\label{tab:dimgap}Bounds on the dimension of lightest
     charged state assuming the theory has U(1) symmetry, as a function of $c$ and the number $n_D$ of derivatives used in the bootstrap functionals. 
}
\end{table}

Similarly to the results of \cite{Collier:2016cls}, extrapolating in $n_{D}$ and then $c$ provides a parametrically stronger bound than the finite $n_{D}$ analysis.
\begin{equation}\label{bound2}
\Delta_{*} = \frac{c}{\alpha} + \mathcal{O}(1)\,, \ \ \  \alpha > 8\,.
\end{equation}

This bound (\ref{bound2}) is similar to the bounds on the non-charged
state found in \cite{Friedan:2013cba,Collier:2016cls}, though quantitatively different.  The bound in
\cite{Friedan:2013cba} is parametrically weaker, which is not
surprising since that analysis did not use the spins of the
characters, and did not perform any extrapolation in the number of
derivatives.  The bound in \cite{Collier:2016cls} is more analogous
since spins and extrapolations were used; €" the result there is very
slightly stronger ($\alpha \sim 9$) than (\ref{bound2}) for the
charged spectrum.  

\subsection{Bound on Charge-to-Mass Ratio}
\label{sec:boundChargeMassRatio}

In this section, we will present results that there must be a state in the theory with a charge-to-mass ratio 
\be
r \equiv \frac{Q c}{12 \Delta} = \frac{Q}{8 G_N m}, 
\ee
above some critical value $r_*$,
whose value we will determine numerically.\footnote{The value of $r^*$
  increases as the number $n_D$ of derivatives used increases and the
  numeric accuracy improves, though we emphasize that even for low
  number of derivatives the values of $r_*$ are a valid bound proving
  that a state must exist in the theory with
  $\frac{Q}{8 G_N m} > r_*$.}

We try to find the linear functional that\footnote{We also impose a
  dimension cutoff $\rho(\Delta,Q) \geq 0,~~~ \Delta>\frac{100c}{12}$,
  because we want the constraint to be a little stronger that not only
  a state which saturates the ratio bound exists but also the state
  must have finite dimension. Different dimension cutoffs do not
  result in significantly different functionals and bounds. It just
  helps the algorithm to find a functional  that only is negative at
  finite $\Delta$.  }
\begin{align}
  \rho({\rm vac}) \geq 0,& \nn \\
  \rho(\Delta,Q) \geq 0,&~~~ |Q| \leq Q_{\Delta} \equiv \frac{12 r \Delta}{c}~.
  \label{eq:Rcond}
\end{align}
We drop the spin index $l$ by only taking functionals of the form
\begin{align}
  \rho \equiv \sum_{m+k/2 {\rm ~odd,~} k {\rm ~even}} (\partial_t
  + \partial_{\bar t})^m \partial_w^k ~.
\end{align}
For the functional to be positive in a bounded region we make a
change of variables of the form 
\begin{align}
  Q^2 = \frac{\tilde{Q}^2  Q_{\Delta}^2}{\tilde{Q}^2 +1 }~.
\end{align}
$|\tilde Q| \geq 0$ means $|Q| \leq  Q_{\Delta}$, inspired by \cite{Collier:2016cls}.

\begin{figure}[t!]
  \centering

  \includegraphics[width=0.79\textwidth]{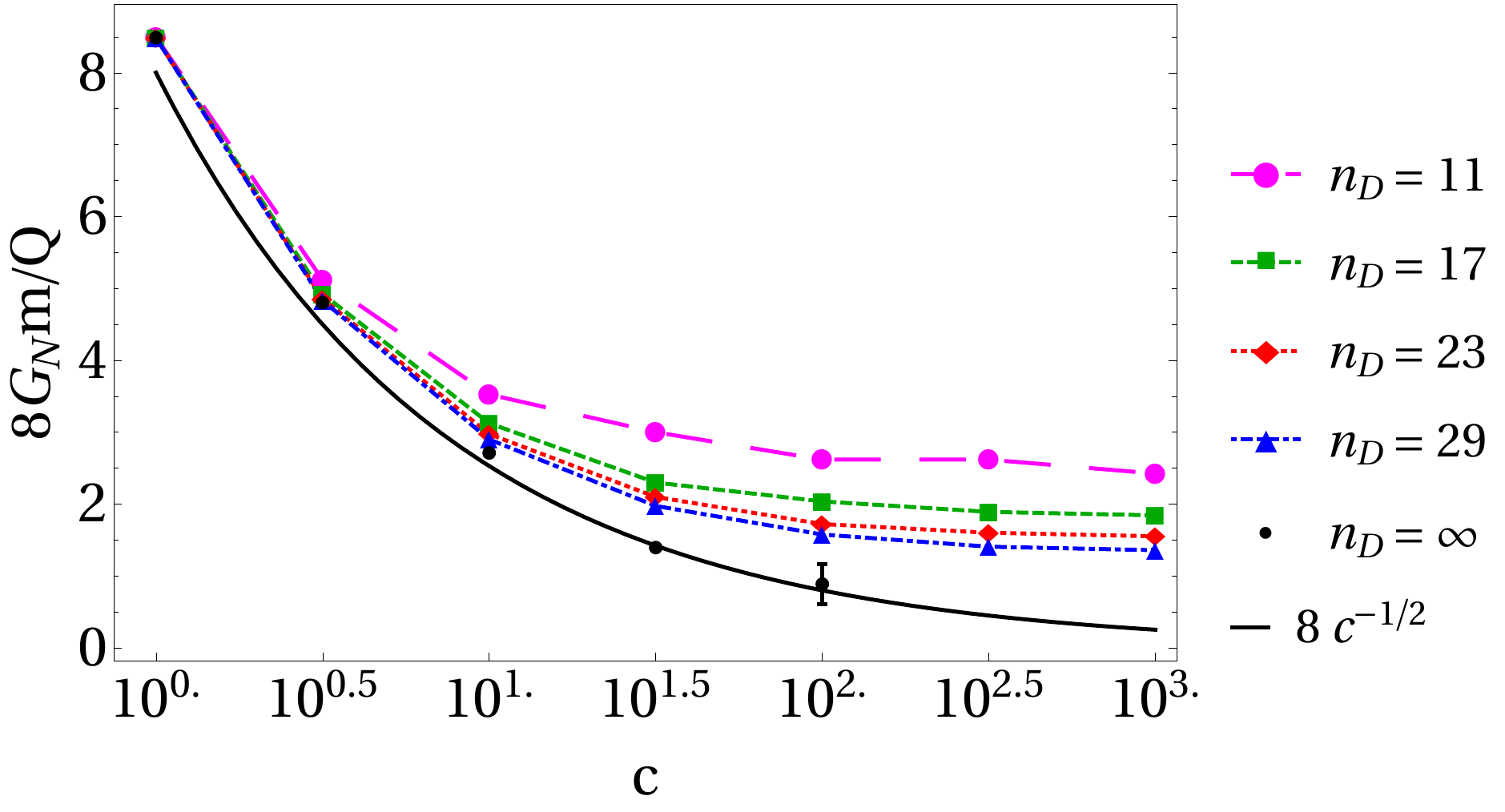}
  \caption{Bound of mass-to-charge ratio as a function of $c$; a trend line $\propto c^{-1/2}$ is shown for comparison. The extrapolation (``$n_D=\infty$'' points) and error bars are computed by performing a fit as a function of $n_D$  and extrapolating to $n_D\rightarrow \infty$ as described in the text. 
  }
  \label{fig:QMbound}
\end{figure}

\begin{figure}[ht!]
  \begin{multicols}{2}
  \begin{tabular}{cc}
$n_D$ & $ \frac{8 G_N m}{Q} $ \\
\hline
27 & 1.38475 \\
29 & 1.31208 \\
31 & 1.29155 \\
33 & 1.24427 \\
35 & 1.19902 \\
37 & 1.16861 \\
39 & 1.14637 \\
41 & 1.10773
\end{tabular}

\hspace{-2in}
  \includegraphics[width=0.6\textwidth]{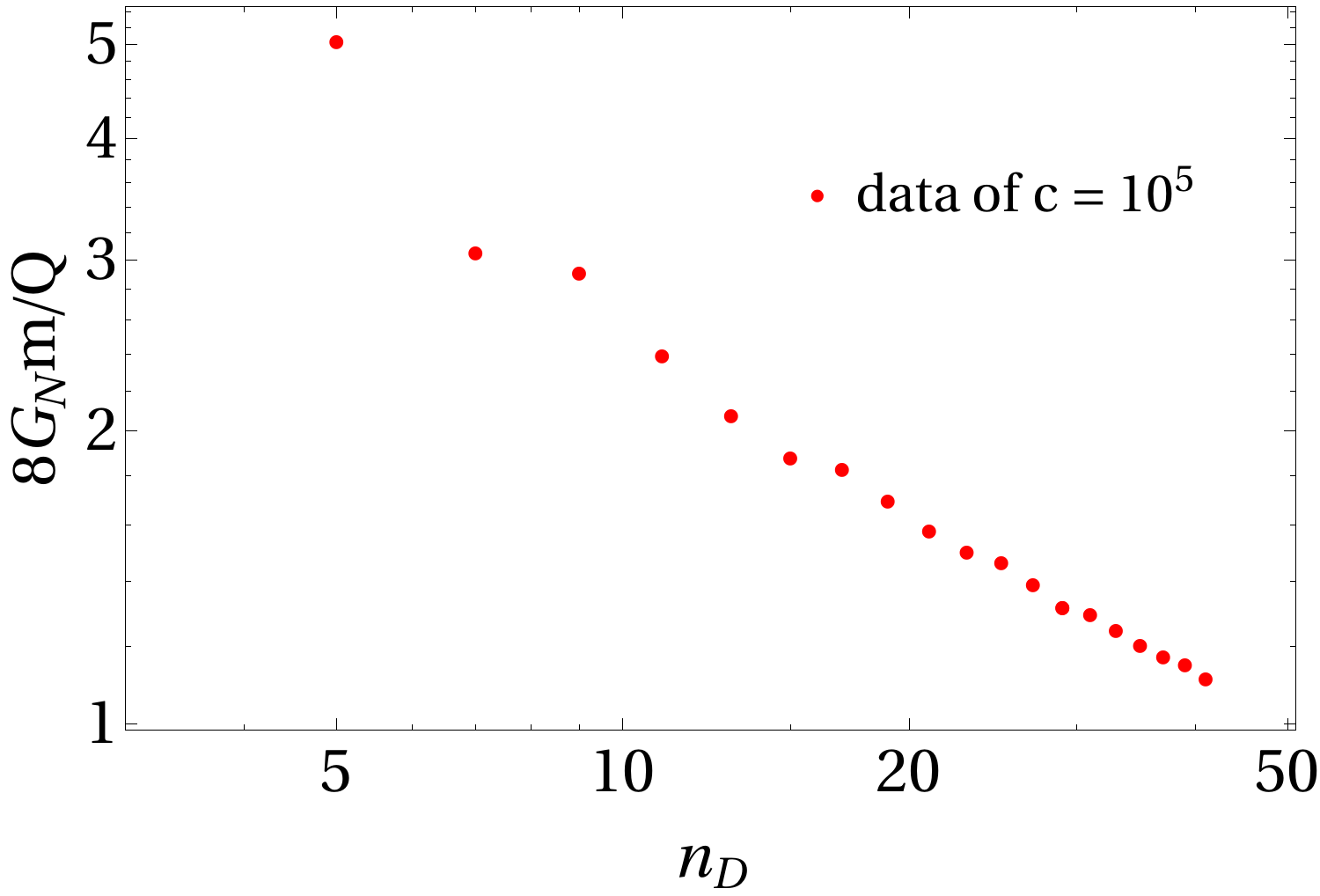}
  \end{multicols}
  \caption{ Upper bound on $\frac{8 G_N m}{Q}$ (that is, there exists
    a state below the bound) as the number $n_D$ of derivatives used
    in the semidefinite programming analysis increases, for the
    specific case $c=10^5$. The bound value is still changing rapidly
    at $n_D = 41$.
  }
  \label{fig:MassQTrend}
\end{figure}

First, we show in Fig. \ref{fig:QMbound} the bound on $r_*$ as a
function of $c$. By inspection, one can see that the larger $c$ is, the longer it takes for the bounds to converge.
To see how the bound depends on the number $n_D$ of derivatives in more detail, in Fig. \ref{fig:MassQTrend}
we focus on a specific
value of $c$, $c=10^5$ and show the resulting bound on $r$ as a function of the number $n_D$ of
derivatives allows in the functional $\rho$. 
The best fit as power law suggests that the optimal
bound on $r^*$ might be significantly better, i.e. $(r^{*})^{-1} \ll 1 $. For comparison, the result in
\cite{Benjamin:2016fhe} was $\left(\frac{8 G_N m}{Q}\right)_* = (r_*)^{-1}  < 4 \sqrt{\pi} = 7.1$.

 In \cite{Benjamin:2016fhe}, it was also shown that even with a small number $n_D$ of derivatives, one could obtain a bound on $\frac{\Delta}{Q}$ at large $c$ that scaled like $\sim c $.  There is an intriguing possibility however that the true bound scales like $c^{1/2}$, and that this is obscured because it takes more and more derivatives to reach this optimal bound as $c$ increases.  The basic idea for why one might expect a $c^{1/2}$ scaling is that in higher dimensions, the scaling of the WGC limit can easily be read off by demanding that the binding energy from gravity and a Coulomb force cancel each other out.  In the AdS$_3$ case, one can think of the binding energy from gravity as $ \frac{3\Delta^2}{c}$, whereas from a $k=1$ Chern-Simons gauge field exchange it is $Q^2$; demanding equality would set $\frac{\Delta}{Q} \approx \sqrt{c/3}$, i.e. $\frac{8 G_N m}{Q} \approx 6.9 c^{-1/2}$. \footnote{One can read off the coefficients by looking at the vacuum conformal block for Virasoro and Kac-Moody algebras in the limit $z\sim 1$ \cite{Fitzpatrick:2014vua,Fitzpatrick:2015zha}.}
 
For comparison, in Fig.  \ref{fig:QMbound} we have also shown a trend line at $c^{-1/2}$, which becomes further below our best numeric bound  with $n_D=29$ derivatives as $c$ increases.  We can try to estimate the optimal bound by taking our result at each $c$ as a function of $n_D$ and extrapolating to $n_D = \infty$. The ``$n_D=\infty$'' points we  show in Fig. \ref{fig:QMbound} fit our results starting at $n_D \ge 15$  in order to get the extrapolation. However, there is significant uncertainty in the resulting estimate, as can be gauged by the fact that performing the fit starting at smaller or larger values of $n_D$ gives different answers.  In Fig. \ref{fig:MassQTrend} we have shown the bound as a function of $n_D$, where one can explicitly see that the bound is still changing rapidly as a function of $n_D$ even at the upper range of what we have been able to achieve numerically. In Fig.  \ref{fig:QMbound}, the error bars we have shown indicate the range over all the different positive values  we obtain if we perform the fit over $n_D \ge 11, n_D \ge 13, \dots n_D \ge 19$.  
  With better numerical accuracy at large values of $c$, it should be possible to more firmly establish this scaling behavior.

\subsection{Bound on Lowest Charge}
\label{sec:ChargeBound}

Next we will focus on bounds on the  lowest
charge $Q_*$ of all charged states in the theory.  We will first consider the charge $Q$ only, and then see how to do better by including information on dimensions and spins.

To determine the upper bound on the gap to the smallest $|Q|$ of all the charged states, we want to find a linear functional satisfying the following conditions: 
\begin{align}
  \rho({\rm vac}) \geq 0,& \nn \\
  \rho(\Delta,0) \geq 0,&~~~ \Delta \geq  0 \nn \\
  \rho(\Delta,Q) \geq 0,&~~~ \Delta \geq 0 {\rm~~~And~~~} |Q| \geq
                          Q_*
\end{align}

\begin{figure}[ht]
  \centering
  \includegraphics[width=0.6\textwidth]{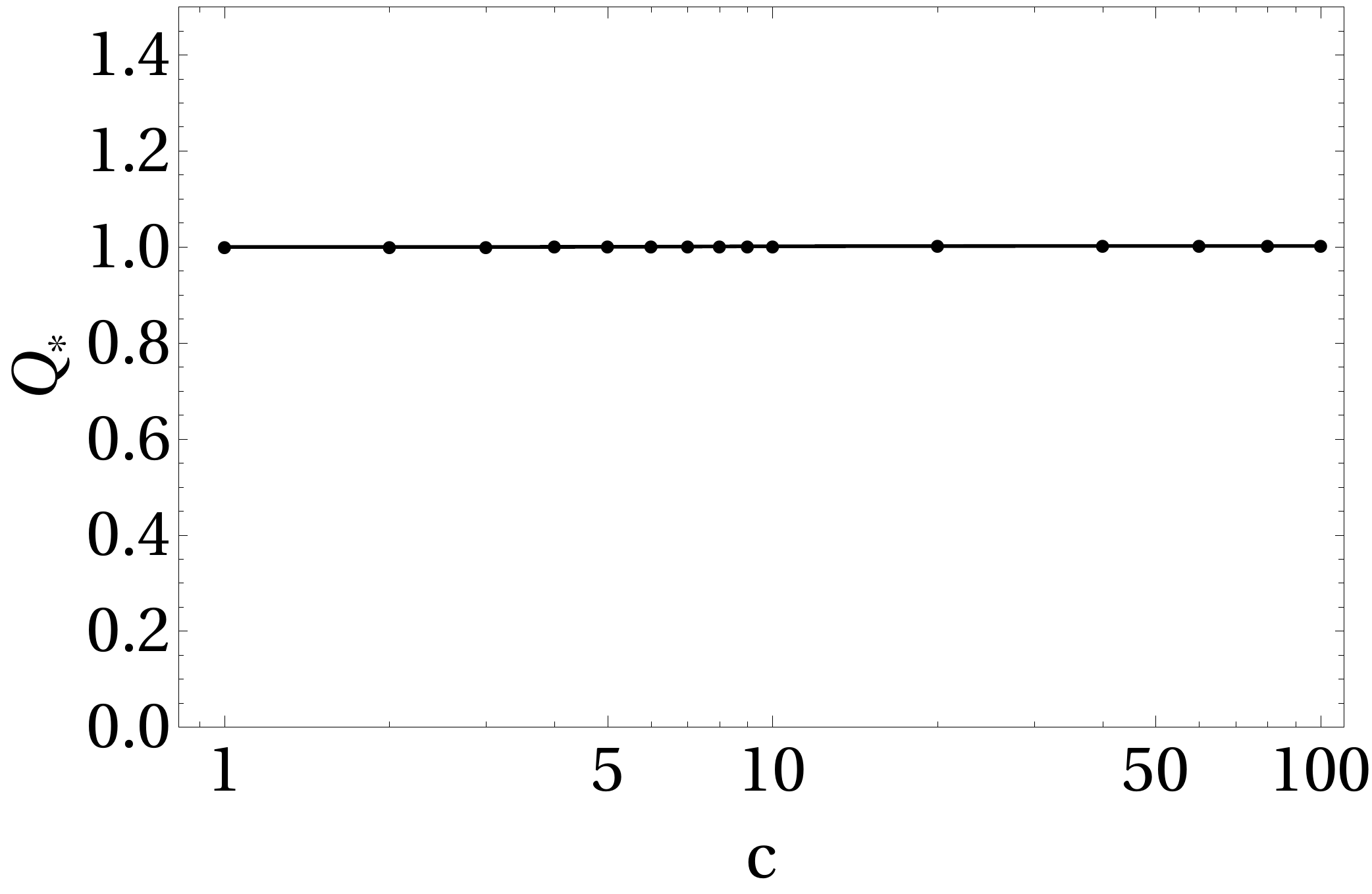}
  \caption{\label{fig:qStarAgainstC} We obtain an upper bound, shown here, on the smallest nonzero charge $Q_*$; the bound is
    $Q_* \le 1$  for all $c$. }
\end{figure}

The resulting bound on $Q_*$ is shown in Fig.~\ref{fig:qStarAgainstC}. The result
is somewhat surprisingly always just $Q_* = 1$. This may be because in some theories with $c\leq 1$ a state of $Q=1$ saturates
the bound and theories of larger $c$ can be constructed as a direct
product of such theories and other algebras.

In any case, we next turn to including information on dimensions by  bounding gaps in  both $Q$ and $\Delta$ simultaneously -- $Q_*$ as the lowest charge of charged
states and $\Delta_*$ as the lowest dimension of all non-vacuum states.  To obtain such a bound, the linear functional $\rho$ should satisfy
\begin{align}
  \rho({\rm vac}) \geq 0,& \nn \\
  \rho(\Delta,0) \geq 0,&~~~ \Delta \geq  \Delta_* \nn \\
  \rho(\Delta,Q) \geq 0,&~~~ \Delta \geq \Delta_* {\rm~~~And~~~} |Q| \geq
                           Q_*
\end{align}
We take the linear functional to have no
spin information.

\begin{figure}[ht!]
  \centering
  \includegraphics[height=0.3\textwidth]{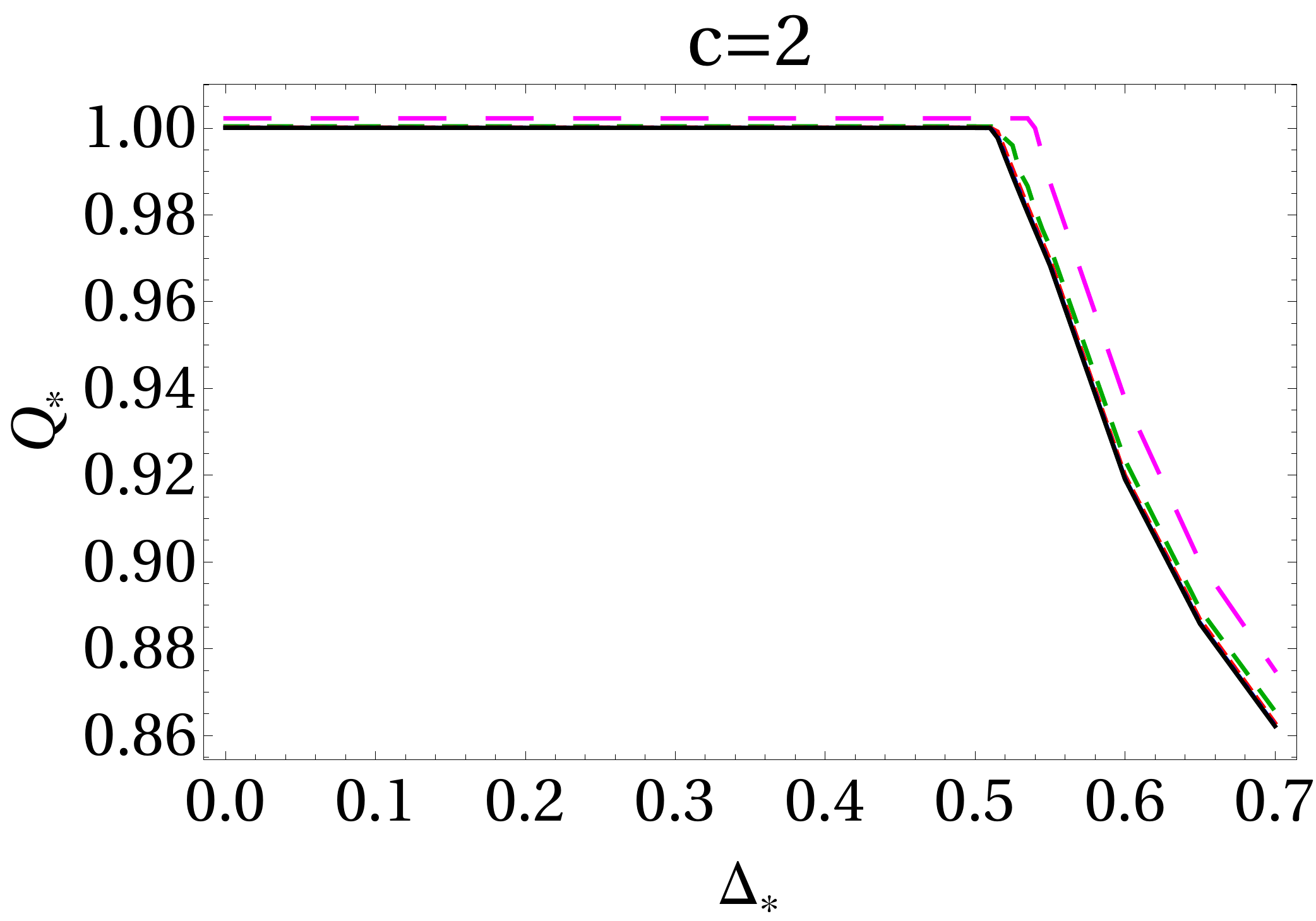}~
  \includegraphics[height=0.3\textwidth]{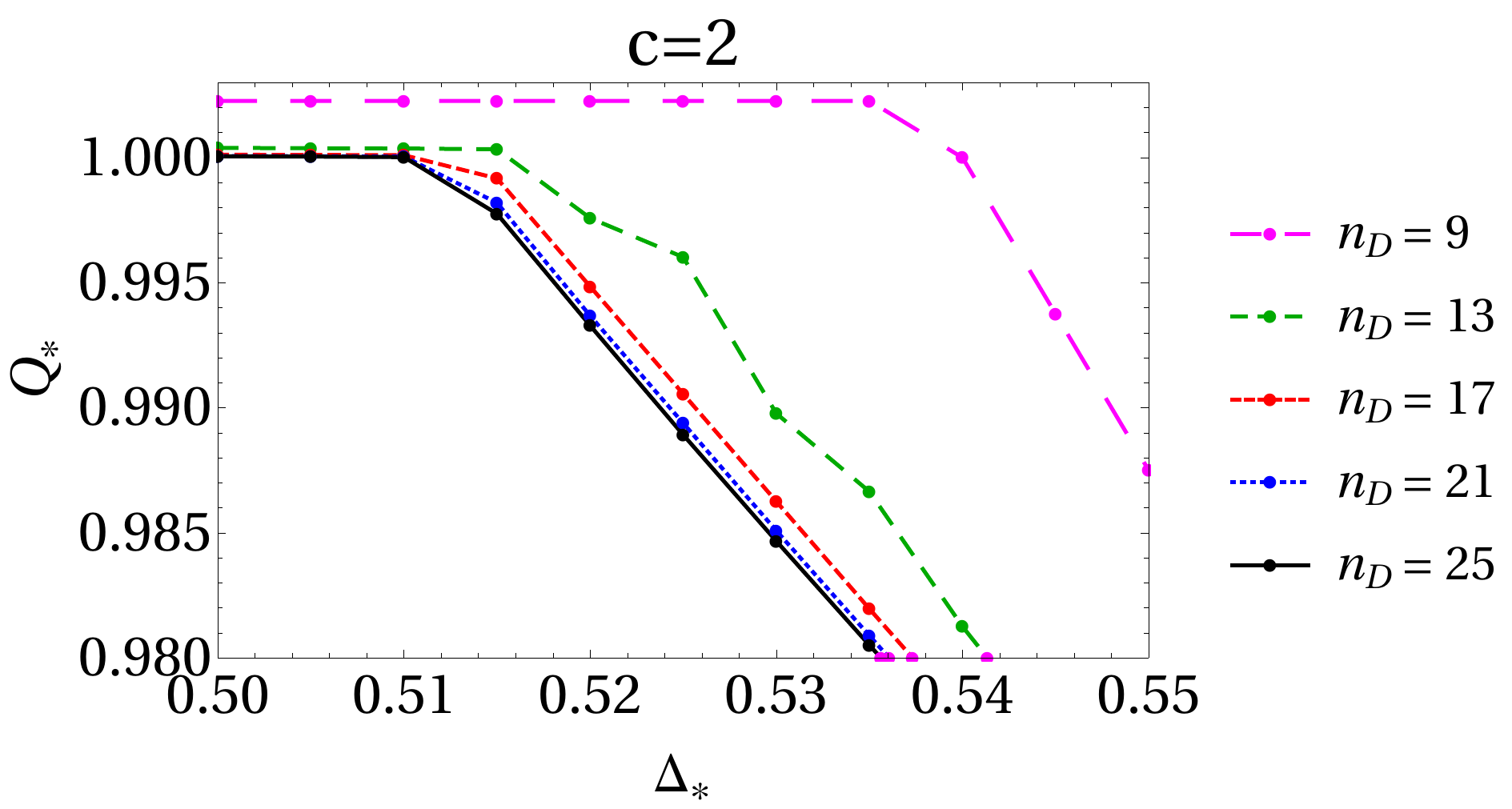}~
  \caption{Bound on the charge gap $Q_*$ and the scalar dimension gap
    $\Delta_*$ at $c=2$.  The region near the kink in the left plot is magnified and shown in the right plot.}
    \label{fig:figBothGapU1_c2_bounds}
  \end{figure}

At each individual $c$, the bound carves out a region in a two-dimensional parameter space.  The exclusion curve of an $c=2$ example is shown in
Fig.~\ref{fig:figBothGapU1_c2_bounds}. There is a kink at $\Delta_* \approx
0.5$ which has a bound $Q_* \leq 1$. It would be interesting to identify what if any theory lives at this kink.

Finally, the simutaneous $\Delta_*$ and $Q_*$ approach can be even more
powerful if we turn on spin. For this case the example we choose is
$c=8$. In \cite{Collier:2016cls} the Sugawara theory $E_8$ lattice
shows up as a kink of bounds on lowest dimension of scalar
primaries. 
We seek a linear functional $\rho$ satisfying
\begin{align}
  \rho({\rm vac}) \geq 0,& \nn \\
  \rho_0(\Delta,0) \geq 0,&~~~ \Delta \geq  \Delta_* \nn \\
  \rho_l(\Delta,0) \geq 0,&~~~ \Delta \geq |l| \nn \\
  \rho_0(\Delta,Q) \geq 0,&~~~ \Delta \geq \Delta_* {\rm~~~And~~~} |Q| \geq
                           Q_* \nn \\
  \rho_l(\Delta,Q) \geq 0,&~~~ \Delta \geq |l| {\rm~~~And~~~} |Q| \geq
                            Q_*  
\end{align}
for all $l \in \mathbb{Z}_{\geq 0}$.

\begin{figure}[ht!]
  \centering
  \includegraphics[width=0.6\textwidth]{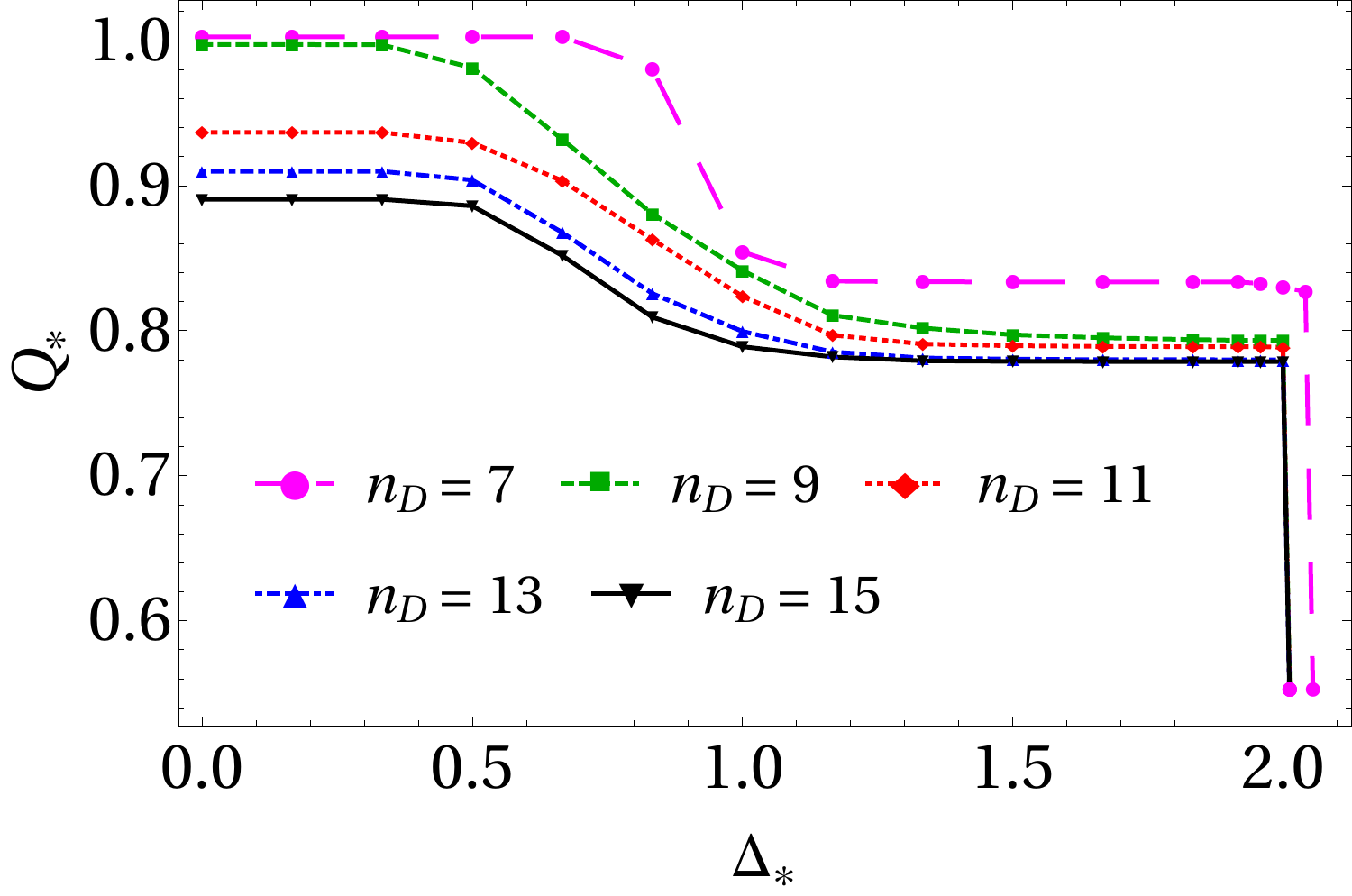}
  \caption{Bound on the charge gap $Q_*$ and the scalar dimension gap
    $\Delta_*$ at $c=8$.}
    \label{fig:QGapSimultaneous_c8}
\end{figure}
The two parameter plot of $\Delta_*$ and $Q_*$ is shown in
Fig.~\ref{fig:QGapSimultaneous_c8}.  Note that we find that $Q_*$ at
$\Delta_*=0$ is smaller than 1, better than the bound obtained without
$Q_*$ information\footnote{It is also possible that by assuming
  integer spins we throw away the theory that saturates the $Q_{*}=1$
  bound.}. More interestingly, we see a sharp kink at $\Delta_{*}=2$. In the next
subsection we will analyze the extremal functional of this kink and
see that this kink is the $E_8$ lattice CFT, and we will obtain the dimension and charge spectrum of
the low lying states.

\subsection{Extremal Functional Analysis}

In this subsection, we will use extremal functional analyses to determine the partition function saturating various bounds. 

\subsubsection{Maximal $r_*$ at $c=1$}

\begin{figure}[ht!]
  \centering
  \includegraphics[width=0.6\textwidth]{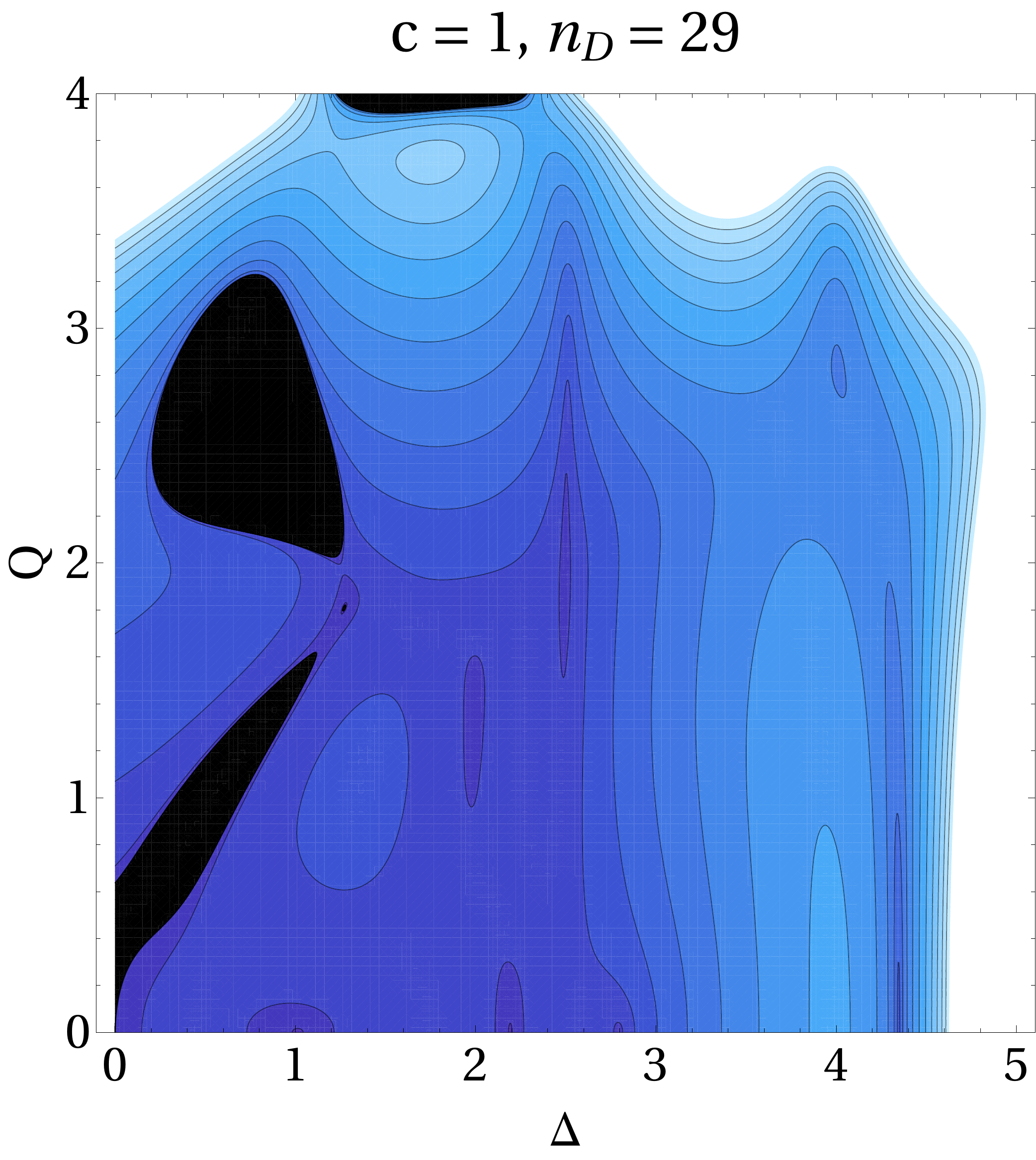}
  \caption{Projection functional $\rho$ of $c=1$ as a function of
    $\Delta$ and $Q$ computed at $N=29$. The black regions are where $\rho \le 0$; according to our criterion (\ref{eq:Rcond}), $\rho$ is $\ge 0$ in the allowed region $Q\le \frac{12 r_* \Delta}{c}$, so the extremal spectrum comprises states where $\rho=0$ in this region. There is a line of small black dots where $\rho=0$ along the $Q=0$ axis that are difficult to see and so we plot $\rho$ along this line in Fig. \ref{fig:restrictedExtremalFunctional_c_1}.}
    \label{fig:ProjectiveRho_c_1}
\end{figure}

Our first application of extremal function methods   will be to the bound on the charge-to-mass ratio $r$. Since our bounds have  converged for $c=1$ and we can 
consider the extremal
functional $\rho$ for this case; by design, $\rho$ is non-negative on the space of states we allow, and so the states in the theory must be at the places where $\rho$ vanishes. The functional depends on both $\Delta$ and $Q$, so
the extremal spectrum contains more data as shown by
Fig.~\ref{fig:ProjectiveRho_c_1}. 

\begin{figure}[ht!]
  \centering
  \includegraphics[width=0.5\textwidth]{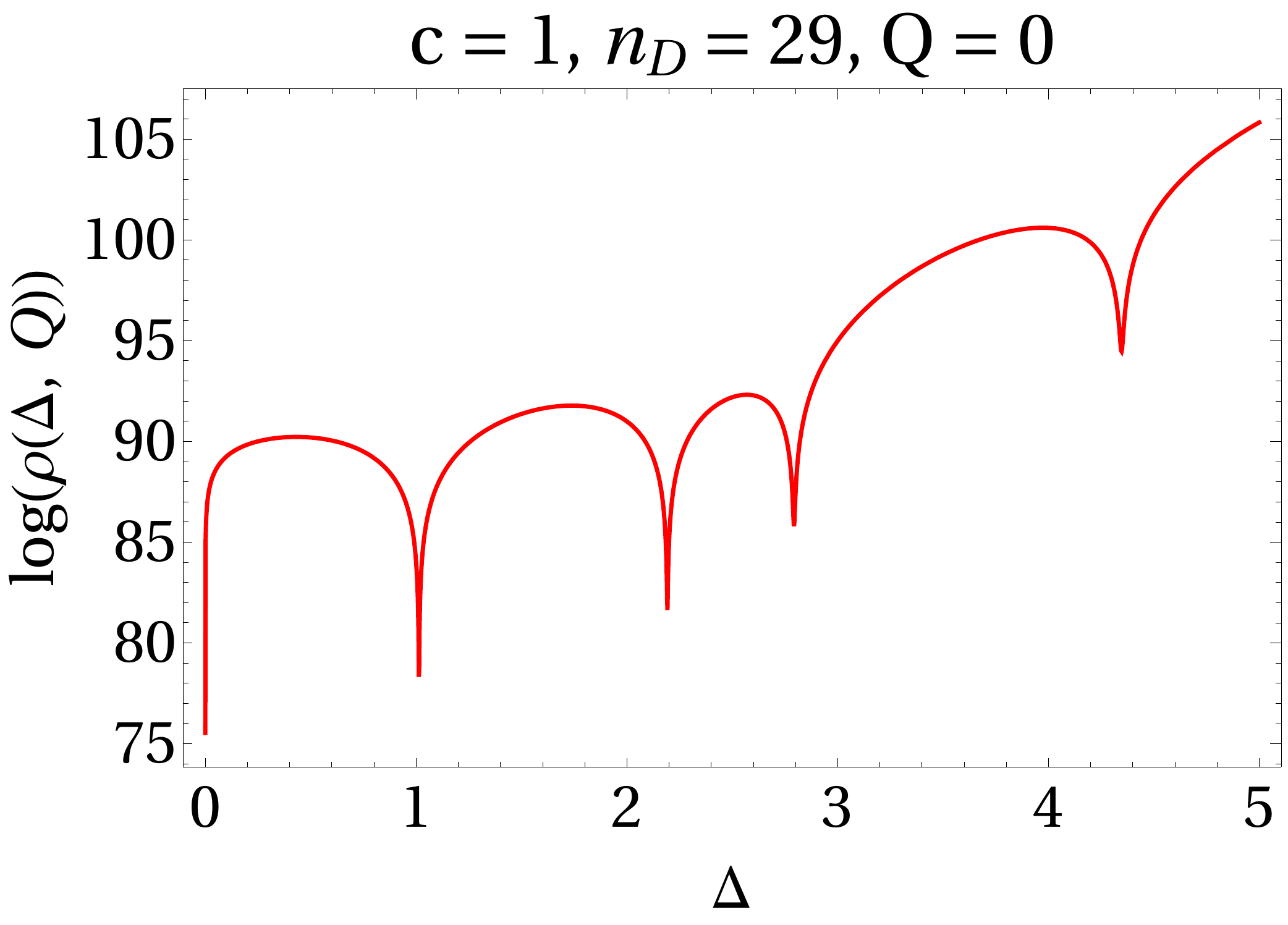}~
  \includegraphics[width=0.5\textwidth]{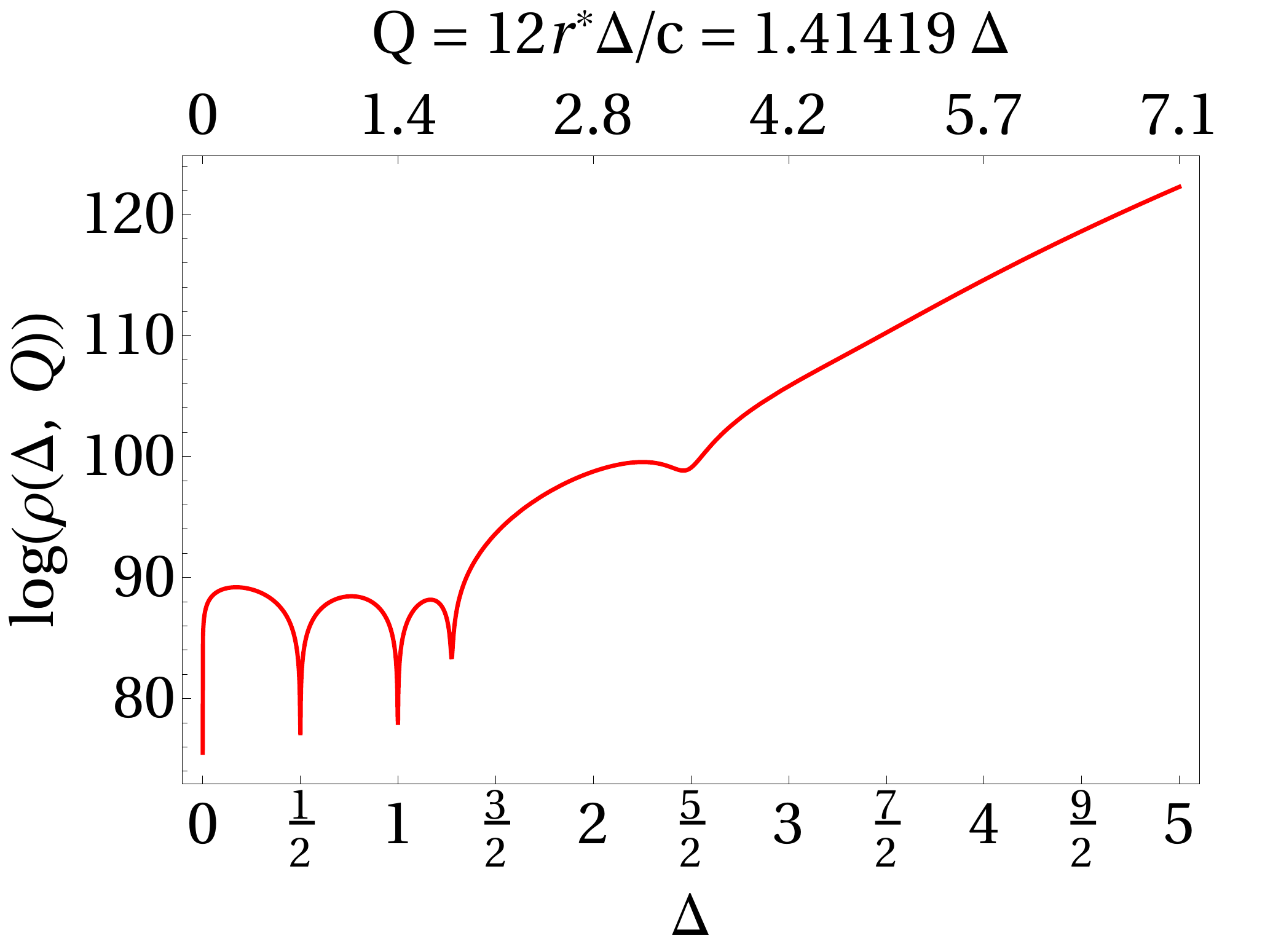}
  \caption{ The functional restricted at $Q=0$ is shown on the
    left. The states that saturate the best bound on
    $\frac{8 G_N m}{Q}$ is shown on the right.  }
    \label{fig:restrictedExtremalFunctional_c_1}
\end{figure}

The zeros of $\rho$ occur at points and can be difficult to see in Fig. \ref{fig:ProjectiveRho_c_1}. In Fig. \ref{fig:restrictedExtremalFunctional_c_1}, we show $\rho$ along two particularly relevant lines: the neutral ($Q=0$) states, and the states that saturate the mass-to-charge ratio, i.e. $Q= \frac{12 r_* \Delta}{c}$.

\subsubsection{Sequential $\Delta_*$ and $Q_*$ Approach - Revisiting the $E_8$  Lattice}

In Sec.~\ref{sec:ChargeBound} we found a kink in the simultaneous
$\Delta_*$ and $Q_*$ approach with spin information at $c=8$. In order to show that the kink is
indeed the level 1 $E_8$ lattice, we can look for extremal flavored
partition functions by using a ``two-step'' approach where we first
solve for the spectrum of dimensions and then solve for the spectrum
of charges.  The idea is that we can use the extremal functional
method on the unflavored partition function, maximizing the gap in
dimensions of operators.  This step is just the standard extremal
functional method and it will just reproduce previous results
\cite{Collier:2016cls}.  Then, having fixed the weights of the states
in the theory, we can impose a gap in the charge of the states in the
theory, allowing only the weights $(h,\bar{h})$ found previously.

\begin{figure}[t!]
  \centering
  \includegraphics[width=0.9\linewidth]{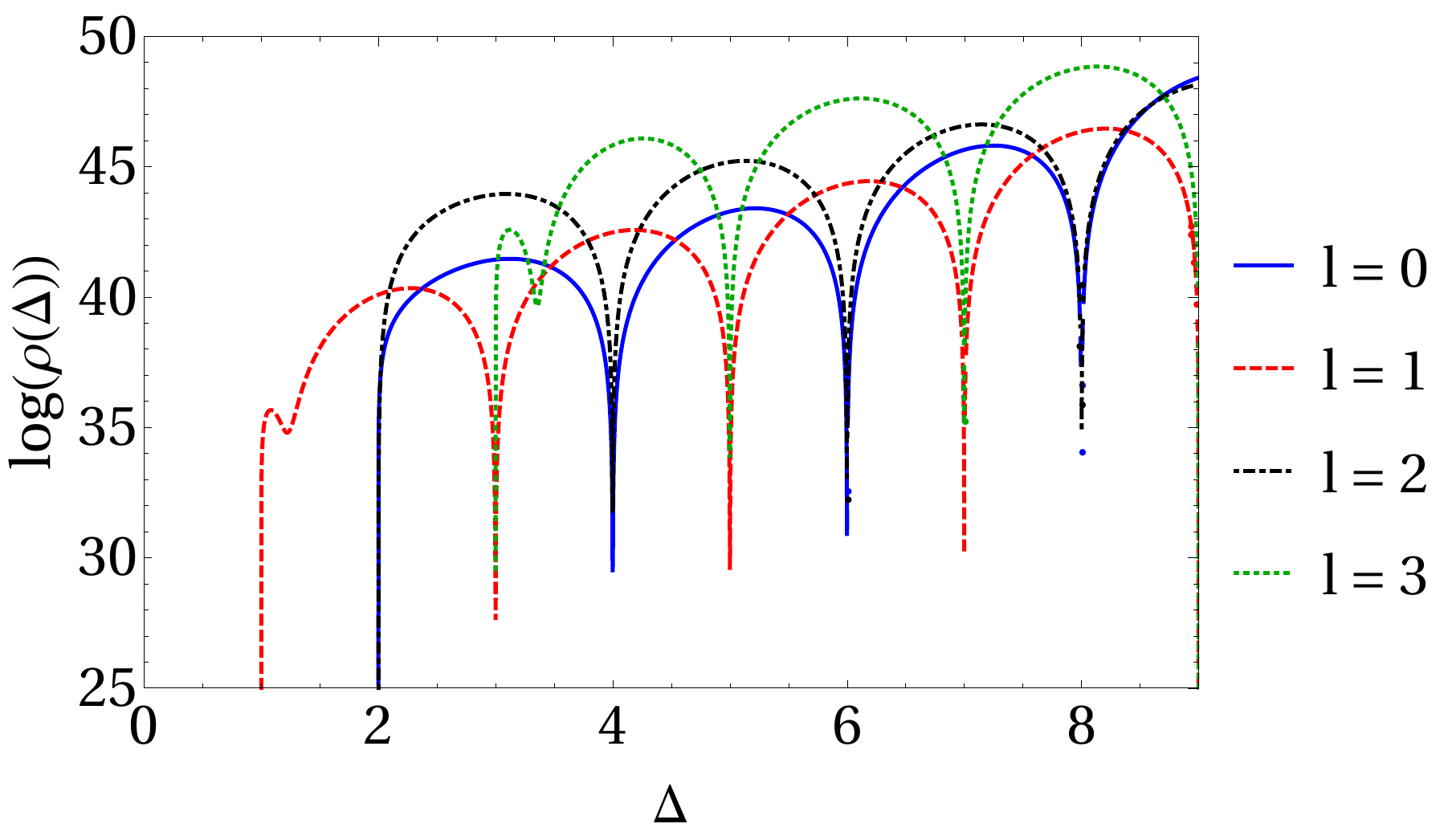}
  \caption{Extremal functional of $c=8$ theory, at $n_D=35$.} 
  \label{fig:fig8extr}
\end{figure}

For concreteness, we will focus on the case $c=8$ as a representative example.  As shown in \cite{Collier:2016cls}, the gap in dimensions is maximized at $\Delta_*=2$ by the $E_8$ theory at $k=1$ (this theory can be described as 8 free bosons on an $E_8$ lattice), and the extremal functional method allows one to independently derive the partition function of this theory.  We have reproduced the extremal functionals $\rho_\ell(\Delta)$ at each spin $\ell$ in Fig.~\ref{fig:fig8extr}, which implies the spectrum is
\begin{align}
  \begin{cases}
    \Delta = 2,4,6,\ldots,& l=0,\\
    \Delta = l,l+2,l+4,\ldots,& l\neq 0 .\\
  \end{cases} 
\end{align}

\begin{figure}[ht!]
  \centering
  \includegraphics[width=0.9\linewidth]{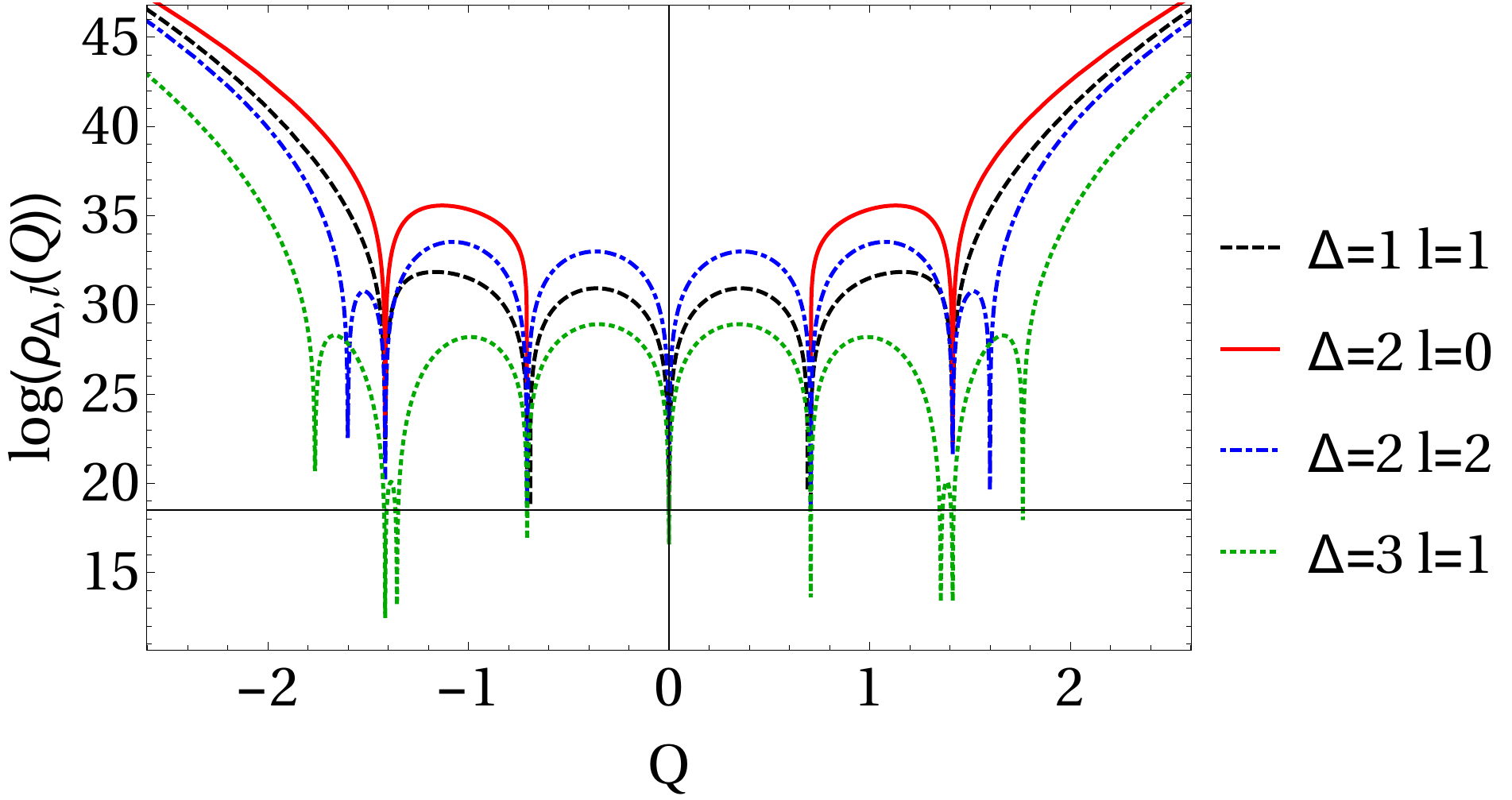}
  \caption{
    Extremal functional of $Q$ at $n_D=19$ when the gap on $Q$ is maximized
    at $\frac{1}{\sqrt{2}}$.}
    \label{fig:c8ExtremalQ}
\end{figure}

Moving on to the spectrum of charges, we find that the gap  $Q_*$ is maximized at $Q_* = \frac{1}{\sqrt{2}}$.  The corresponding extremal functionals $\rho_{\Delta, \ell}(Q)$ at each dimension $\Delta$ and spin $\ell$ are plotted in Fig.~\ref{fig:c8ExtremalQ}.  We note that this is  not the flavored partition function that one obtains if one turns on a chemical potential for the charge $J = \partial \phi$ in the $E_8$ lattice description; that choice corresponds to the spectrum of charges $\frac{1}{2}\mathbb{Z}$ rather than $\frac{1}{\sqrt{2}} \mathbb{Z}$.  Instead, if one chooses one of the length-2 vectors $\vec{\alpha}$ in the $E_8$ lattice, then
\be
J \equiv V_{\vec{\alpha}} \equiv \frac{1}{\sqrt{2}} \left( e^{\vec{\alpha}\cdot \vec{\phi}} + e^{-\vec{\alpha} \cdot \vec{\phi}}  \right)
\ee
has $k=1$, and the lowest charged states include for instance $V_{2\vec{\alpha}}$, with charge $\frac{1}{\sqrt{2}}$. 

One can see in Fig.~\ref{fig:c8ExtremalQ} that the extremal functional
has zeros at around $0, \pm \frac{1}{\sqrt{2}}$ and
$\pm \frac{2}{\sqrt{2}}$ for all $\Delta, \ell$, and we expect that
this would continue to be true at $ \frac{n}{\sqrt{2}}$ for all
$n \in \mathbb{Z}$ as the number $n_D$ of derivatives used in the
analysis approaches infinity. 
 Because these dimensions and charges  appear to follow such a simple pattern, we will proceed by 
 assuming this pattern continues. 
Then, with
the allowed weights $\Delta, \ell$ and charges $Q$ fixed in the
theory, solving the modular bootstrap equation reduces to a linear
programming problem, which is much more efficient numerically.\footnote{Furthermore, since this linear programming analysis fixes the partition function for us to be a particular flavoring of the $E_8$ theory, by uniqueness  it will be the correct one, justifying the original Ansatz.  } We
obtain the occupation numbers indicated in
Table~\ref{tab:c8ExtremalQ1}, where we have flavored separately by both holomorphic and anti-holomorphic charges $Q$ and $\bar{Q}$.
 We show the occupation numbers of the
conserved $ \ell  = 1$ currents  assuming the extremal charge spectrum  $Q= \frac{n}{2}$ (right). In addition, it is straightforward to repeat the analysis assuming $Q = \frac{n}{\sqrt{2}}$ (left) for comparison.  In both cases, we obtain a total of $248$ currents each in the
holomorphic and anti-holomorphic part, but distributed differently among different charges in the two cases.

\begin{table}[t!]
\begin{center}
\begin{tabular}{ccccc}
$ \Delta$ & $\ell$ & $|Q|$ & $|\bar{Q}|$ & $d_{\Delta, \ell, Q}$ \\
\hline 
  1 & 1 & 0 & 0 & 134 \\
  1 & 1 & $\frac{1}{\sqrt{2}}$ & 0 & 112 \\
  1 & 1 & $\sqrt{2}$ & 0 & 2 \\
  1 & -1 & 0 & 0 & 134 \\
  1 & -1 & 0 & $\frac{1}{\sqrt{2}}$ & 112 \\
  1 & -1 & 0 & $\sqrt{2}$ & 2 \\
\end{tabular}\hspace{1in}
\begin{tabular}{ccccc}
$ \Delta$ & $\ell$ & $|Q|$ & $|\bar{Q}|$ & $d_{\Delta, \ell, Q}$ \\
\hline 
  1 & 1 & 0 & 0 & 92 \\
  1 & 1 & $\frac{1}{2}$ & 0 & 128 \\
  1 & 1 & 1 & 0 & 28 \\
  1 & -1 & 0 & 0 & 92 \\
  1 & -1 & 0 & $\frac{1}{2}$ & 128 \\
  1 & -1 & 0 & 1 & 28 \\
\end{tabular}
\end{center}
\caption{Occupation numbers from a linear programming analysis.  The left table assumes states at $Q \in \frac{1}{\sqrt{2}} \mathbb{Z}$, whereas the right assumes $Q \in \frac{1}{2} \mathbb{Z}$. 
}
\label{tab:c8ExtremalQ1}
\end{table}

\section{Non-Abelian Bounds}

\subsection{Bounds on gaps in operator dimensions}
\label{sec:NAbounds}

Next we turn to a numeric analysis of gaps in non-abelian theories.  In some cases, the results are somewhat stronger or weaker depending on whether or not we allow for states that saturate the unitarity bound $2 k h \ge Q^2$, which we will refer to as ``extremal states'', and whether or not we impose gaps in all charge sectors.  We will present results starting with the strongest assumptions first. 

In all cases, we will present only the results for $SU(2)$ gauge group at level $k=1$.  We have analyzed larger gauge groups and higher levels and the results are qualitatively similar, though the rate of numeric convergence is worse; some preliminary results for $SU(2)$ with $k=2$ are shown in appendix \ref{app:su2k2}.

\begin{figure}[th!]
  \centering
  \includegraphics[height=0.27\textwidth]{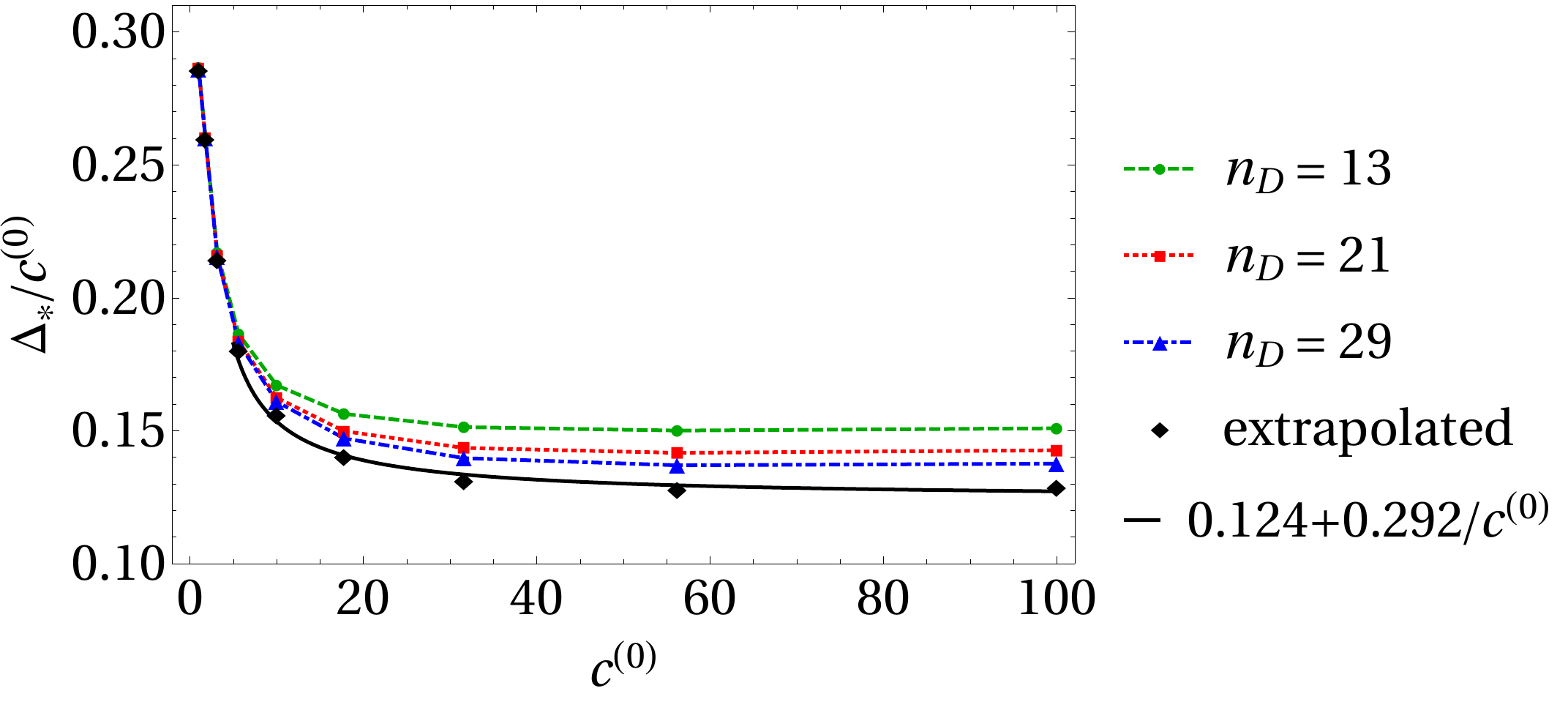}~ \hspace{-.3in}
    \includegraphics[height=0.26\textwidth]{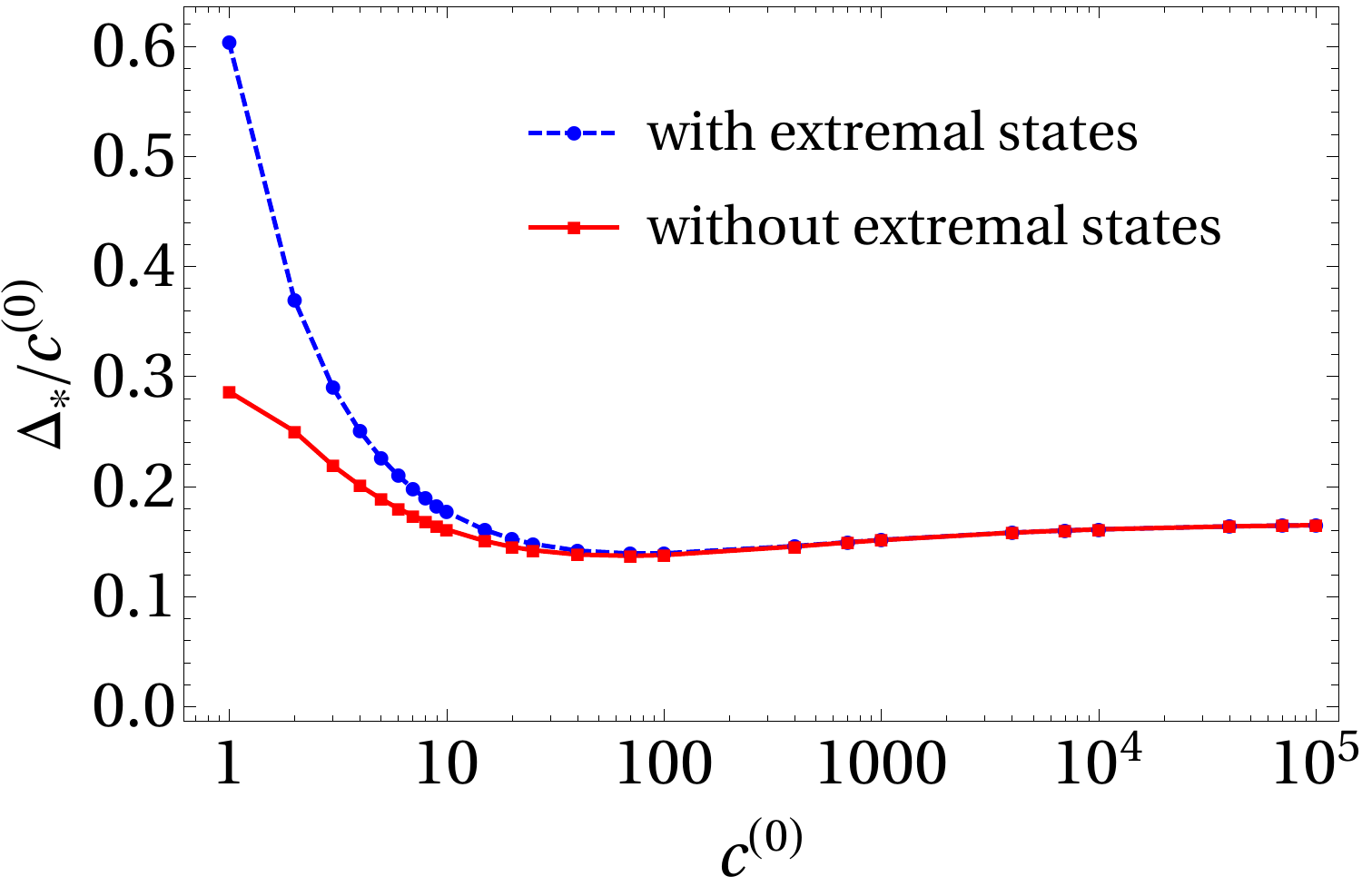}
    \caption{ Bound on SU(2) $k=1$ gaps in $\Delta$ universal for all
      representations for $1 \leq c \leq 100$. {\bf Left:} Bounds
      obtained with different values of $n_D$ when extremal states are
      not allowed. Dashed lines from blue to red are computed data of
      $3 \leq n_D \leq 29$. The solid black line is extrapolated from
      data of $ 11 \leq n_D \leq 29$ using a function linear in
      $1/n_D$. {\bf Right:} Bounds extrapolated to $n_D = \infty$ for
      the analysis without extremal states compared to the result when
      extremal states are allowed.  The difference is negligible at
      large $c$ but significant at small $c$.
    }
  \label{fig:su2extrapolated}
\end{figure}

To begin,  we will set the gap in all representations to be the same and restrict to the partition function at $q=\bar{q}$;
the resulting  gap value will be the lowest dimension of the
primary operators.
This ``uniform bound''  is shown in Fig. \ref{fig:su2extrapolated}.  

We have actually done two slightly different analysis, which are compared to each other on the right in Fig. \ref{fig:su2extrapolated}.  These analyses differ in whether or not we allow states in the non-trivial representations with $h^{(0)}= \bar{h}^{(0)}=0$, which saturate the unitarity bound in both the holomorphic and anti-holomorphic sectors and which we will refer to as ``extremal states;'' in the analysis where such states are included, the ``gap'' for each representation is defined as the smallest $\Delta^{(0)}$ among the non-extremal states.   As one can see, the difference between the results of the two analyses is significant at small $c$ but becomes negligible as $c$ approaches $\infty$. 

Ultimately, this result does not tell us much more than one learns from previous similar analyses without flavored information; all we learn here is that there must be some state in the theory with $\Delta^{(0)}$ below some value, which is very similar to the bound on the same quantity from the unflavored modular bootstrap. 

Next, however, we will turn to an analysis that maximizes the bounds separately in different sectors, and this is where we will start to find something qualitatively new compared with  what is possible with the unflavored modular bootstrap.

\begin{figure}[htbp]
  \centering
  $$
  \begin{array}{cc}
    \hspace{-.3in}
    \includegraphics[height=0.27\linewidth]{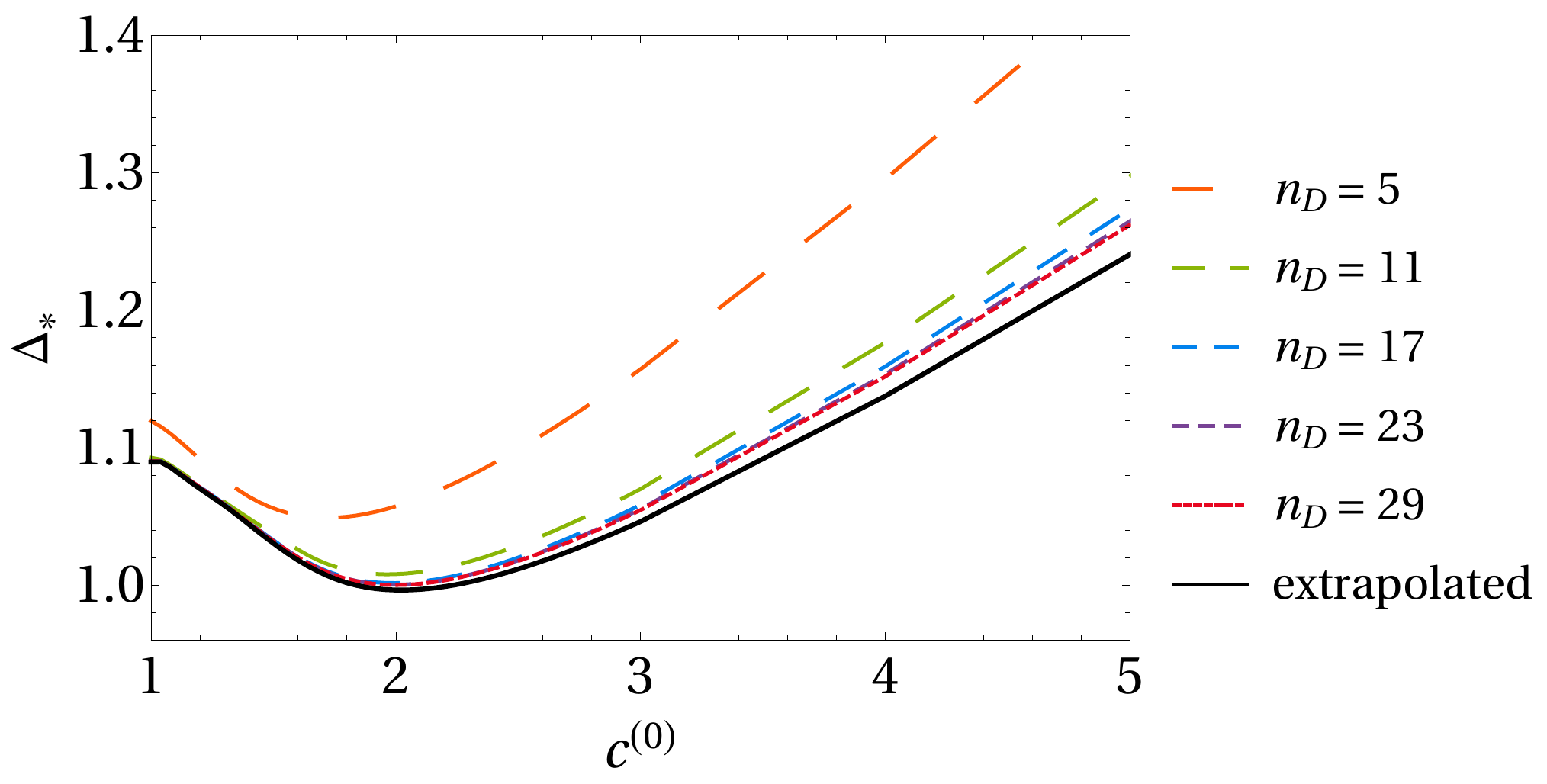}
    &\raisebox{0.01in}{
      \includegraphics[width=0.4\linewidth]{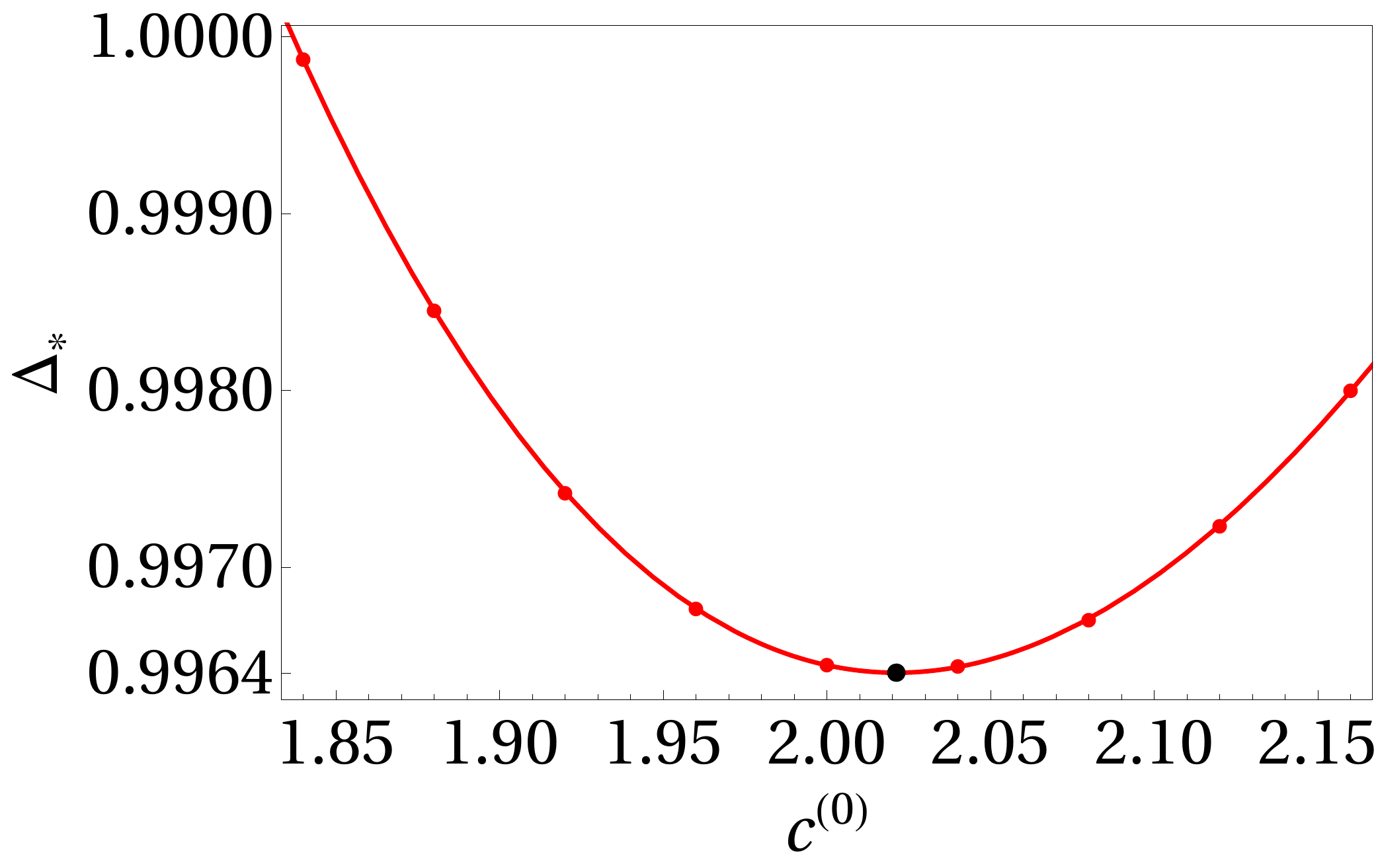}
    }\\
  \end{array}
  $$
  \caption{{\bf Left:} The upper bound on the gap in the dimension of
    primaries, $\Delta_{*}$, in the trivial representation
    obtained at increasing derivative order of the linear functional
    (from blue to red, up to $n_D = 29$). The black curve is the
    extrapolated value. {\bf Right:} Blown-up plot of the-near minimal
    $\Delta_{*}$ region. The minimal is
    expected at $2.00\leq c \leq 2.04$,
    $\Delta_{*} \approx 0.995$.
    }
    \label{fig:su2specificbound}
\end{figure}

In particular, 
we will maximize the gap in the trivial representation, and not impose any constraint on the gaps in the other representations. In equations, our conditions are
\begin{equation}
  \rho_{\lambda,\bar \lambda}(\Delta^{(0)})\geq 0 {\rm ~when~}
  \begin{cases}
    \Delta^{(0)} \geq \Delta_{*},&\lambda=(0){\rm~and~} \bar{\lambda}=(0),\\
    \Delta^{(0)} \geq 0,&\lambda {\rm~or~} \bar{\lambda}\neq (0),\\
  \end{cases}
\end{equation}
in SU(2) $k=1$, weight $\lambda$ (or $\bar \lambda$) takes values $(0)$ or $(\frac{1}{2})$. The resulting bound on the neutral sector gap is shown as a function of $c$ in Fig. \ref{fig:su2specificbound}.\footnote{By contrast with the previous subsection, here we find that the bound is exactly the same whether or not we allow extremal ($h^{(0)}=\bar{h}^{(0)}=0$) states in the non-trivial representations.}  Notably, there is a minimum of about $\Delta_* =1$ near $c \equiv c^{(0)}+1 =3$. We next turn to a more detailed study of this point.

\subsection{Extremal Functional Analysis at $c=3$}

\subsubsection{Spin-Independent Analysis}

\begin{figure}[ht]
  \centering
  \includegraphics[width=0.7\textwidth]{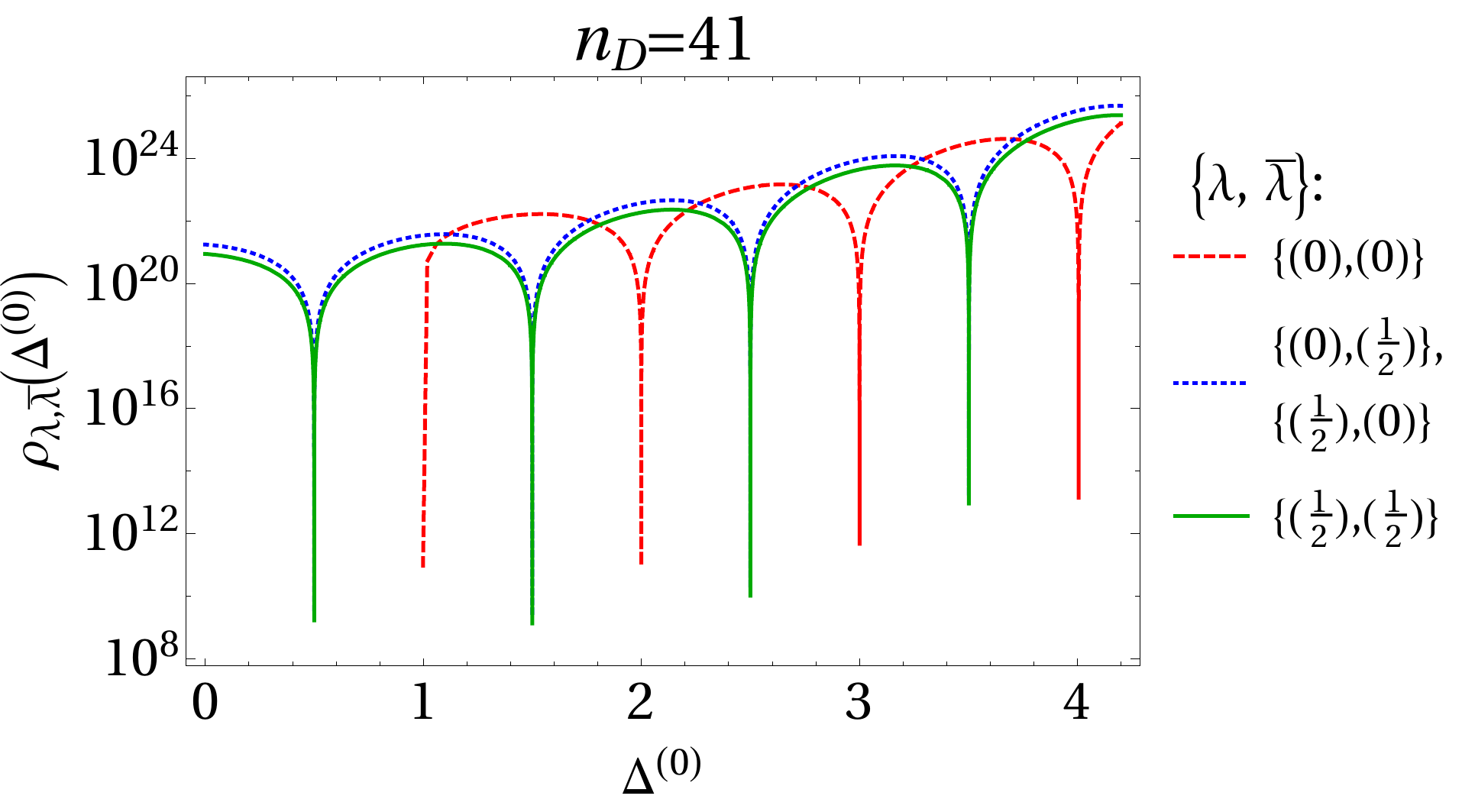}
  \caption{\label{fig:figsu2k1Spectrum}
    Extremal functional of SU(2) $k=1$, $c=3$ spectrum.}
\end{figure}

Our $q=\bar{q}$ analysis in subsection (\ref{sec:NAbounds})
found a minimum gap bound at $c=3$. 
Using the extremal functional method \cite{ElShowk:2012hu},
 the dimensions and the degeneracies of states at this point can be
extracted, with numerical accuracy being best for the lowest dimension
states. The dimensions of states occur at the zeros of the extremal
functional, plotted in Fig.~\ref{fig:figsu2k1Spectrum}.

Furthermore, we find that the occupation numbers of the lowest-energy
states of the partition function are uniquely determined to be
\begin{align}\label{fig:su2specificExtremalPartionFn}
  M_{(0),(0)}(\beta) &\approx \chi_{0}(\beta) + 28\chi_{1}(\beta) + 76\chi_{2}(\beta) +
  274 \chi_{3}(\beta)
  +\ldots
  \\
  M_{(0),(\frac{1}{2})}(\beta) =
  M_{(\frac{1}{2}),(0)}(\beta) &\approx 8\chi_{0.5}(\beta) + 48\chi_{1.5}(\beta)
  +\ldots
  \\
  M_{(\frac{1}{2}),(\frac{1}{2})}(\beta) &\approx 8\chi_{0.5}(\beta) + 48\chi_{1.5}(\beta)
  +\ldots
\end{align}
The subscript on $\chi_{\Delta^{(0)}}$ denotes the non-Sugawara dimension of the state.  At this point, the analysis takes $\tau \equiv \frac{i \beta}{2\pi}$ to be pure imaginary, so no information on spins is used:
\be
\chi_{\Delta^{(0)} } = q^{-\frac{c}{12}} \left\{ \begin{array}{cc} \prod_{n=2}^\infty (1-q^n)^{-2}, & \Delta^{(0)}=0 , \\
 q^{\Delta^{(0)}} \prod_{n=1}^\infty (1-q^n)^{-2}, & \Delta^{(0)} >0 \end{array} \right. 
 \ee

We find that a manifestly modular invariant partition function that reproduces this is
\be
Z(\beta)_{z=\bar{z}} &=& \frac{1}{2 \eta^6 } \left( \left( \theta_3^2 - \theta_4^2 \right) \theta_2^4(\frac{z}{2}) + \left( \theta_2^2 + \theta_4^2 \right) \theta_3^4(\frac{z}{2}) + \left( \theta_3^2 - \theta_2^2 \right) \theta_4^4(\frac{z}{2}) \right) \\
 &=& \frac{1}{2 \eta^6} \left( \theta_2^2 \theta_2^4(\frac{z}{2}) + \theta_3^2 \theta_3^4(\frac{z}{2}) + \theta_4^2 \theta_4^4(\frac{z}{2}) + \left( \theta_3^2 - \theta_2^2 - \theta_4^2 \right) \theta_1^4(\frac{z}{2}) \right) , 
 \label{eq:C3ExtremalPF}
\ee
where $\theta_i\equiv \theta_i(z=0)$.  It is straightforward to check  
that the occupation numbers are non-negative, so that this partition function is unitary, modular invariant, and extremizes the scalar gap. Therefore (\ref{eq:C3ExtremalPF}) is the correct partition function by uniqueness.  

\subsubsection{Spin-Dependent Analysis} 

In the previous subsection, we used extremal functional techniques to determine a unique partition function on the subspace $q=\bar{q}$ when the gap in the scalar sector was maximized for $c=3$.  We can gain much more information about the theory by relaxing the constraint $q=\bar{q}$ and varying $q,\bar{q}$ independently; in particular, the analysis becomes sensitive to the spins $h-\bar{h}$ of the spectrum.  We could continue to use semi-definite programming methods, but they converge less quickly for independent $q,\bar{q}$ than they do for $q=\bar{q}$.  Instead, we can use the fact that we know the spectrum of dimensions from the $q=\bar{q}$ analysis, and the fact that spin is quantized.  This allows us to fix the allowed values of $h,\bar{h}$ to a discrete set, turning the problem into a linear programming problem and thereby making the analysis much more efficient.  

There is a remaining ambiguity, however, which is that we have to make a choice about what spins are allowed.  We find that if we allow only integer spins, there is no allowed partition function and in fact we can reduce the bound on the gap somewhat to about 2/3.  If instead we allow fractional spins, then we find a few different possibilities depending on what spins we allow.  

We will begin with the conventional case where we allow integer and half-integer total spins, $h-\bar{h}$.  The $SU(2)$ Sugawara characters are such that
$M_{(0),(0)}$ only has integer spins, $M_{(0),(1/2)}$ and
$M_{(1/2),(0)}$ only has quarter spins and $M_{(1/2),(1/2)}$ can have
integer and half integer spins. Then to meet the requirement
$M_{(0),(0)}$ and $M_{(1/2),(1/2)}$ can only have spins $\frac{2n}{4}$
and $M_{(0),(1/2)}$ and $M_{(1/2),0}$ can only have spins $\frac{2n+1}{4}$.  Performing the linear programming analysis for such a spectrum (and continuing to maximize the gap in the neutral sector) leads to the following unique set of weights and occupation numbers $d$:\footnote{The reader may notice that the numbers at each dimension in eq. (\ref{fig:su2specificExtremalPartionFn}) do not match the total number of states in the table.  The reason is that without knowledge of spin, there are null states that could not be taken into account in (\ref{fig:su2specificExtremalPartionFn}).  For instance, at level 2, there are a total of 84 states, as compared with 76 in (\ref{fig:su2specificExtremalPartionFn}), because of the 8 conserved currents at spin 1 that consequently have 8 ``null'' descendants at $\Delta= 2$. At $\Delta\ge 3$, the presence of such null states causes states to get reorganized in increasingly complicated ways and it is easiest to check the number of states is the same by constructing the full partition function.}

\begin{center}
\footnotesize
\begin{tabular}{|ccc|}
\hline
$ (\mu, \bar{\mu}) $ & $(\Delta^{(0)} , |\ell^{(0)} |)$ & $d$ \\
\hline 
$(0,0)$ & $(0,0)$ & 1 \\
\hline
 & $(1, 0)$ & 4 \\
 & $(1, \frac{1}{2} ) $ & 16 \\
 & $(1, 1 ) $ &8 \\
 \hline
 & $(2,0)$ & 16 \\
 & $(2, \frac{1}{2} ) $ & 16 \\
 & $(2, 1 ) $ & 24 \\
 & $(2, \frac{3}{2} ) $ & 16 \\
 & $(2, 2 ) $ & 12 \\
 \hline
 & $(3, 0 ) $ & 36 \\
 & $(3, \frac{1}{2} ) $ & 16 \\
 & $(3, 1 ) $ & 48 \\
 & $(3, \frac{3}{2} ) $ & 80 \\
 & $(3, 2 ) $ & 32 \\
 & $(3, \frac{5}{2} ) $ & 48 \\
 & $(3, 3 ) $ & 26 \\
 \hline
 \end{tabular}
 \begin{tabular}{|ccc|}
 \hline
$ (\mu, \bar{\mu}) $ & $(\Delta^{(0)} , |\ell^{(0)} | )$ & $d$ \\
\hline 
 $(\frac{1}{2} , 0 )$  & $(\frac{1}{2}, \frac{1}{4})$ & 8 \\
\hline
 & $(\frac{3}{2}, \frac{1}{4})$ & 0 \\
 & $(\frac{3}{2}, \frac{3}{4})$ & 32 \\
 & $(\frac{3}{2}, \frac{5}{4})$ & 16 \\
 \hline
 & $(\frac{5}{2}, \frac{1}{4})$ & 64 \\
 & $(\frac{5}{2}, \frac{3}{4})$ & 0 \\
 & $(\frac{5}{2}, \frac{7}{4})$ & 56 \\
 & $(\frac{5}{2}, \frac{9}{4})$ & 32 \\
&&\phantom{s}  \\ &&\phantom{s} \\ &&\phantom{s} \\ &&\phantom{s}\\ &&\phantom{s} \\ &&\phantom{s}\\ && \phantom{s}\\ && \phantom{s} \\ 
 \hline
  \end{tabular}
  \begin{tabular}{|ccc|}
  \hline
$ (\mu, \bar{\mu}) $ & $(\Delta^{(0)} , |\ell^{(0)} |)$ & $d$ \\
\hline 
$(\frac{1}{2}, \frac{1}{2}) $  & $(\frac{1}{2}, 0)$ & 4 \\
 & $(\frac{1}{2}, \frac{1}{2})$ & 4 \\
 \hline
 & $(\frac{3}{2}, 0)$ & 16 \\
 & $(\frac{3}{2}, \frac{1}{2})$ & 16 \\
 & $(\frac{3}{2}, 1)$ & 8 \\
 & $(\frac{3}{2},\frac{3}{2}) $ & 12 \\
 \hline
 & $(\frac{5}{2}, 0)$ & 4 \\
 & $(\frac{5}{2}, \frac{1}{2})$ & 48 \\
 & $(\frac{5}{2}, 1)$ & 32 \\
 & $(\frac{5}{2}, \frac{3}{2})$ & 24 \\
 & $(\frac{5}{2}, 2)$ & 40 \\
 & $(\frac{5}{2}, \frac{5}{2})$ & 16 \\
&&\phantom{s} \\ &&\phantom{s} \\  &&\phantom{s}\\ &&\phantom{s} \\ 
 \hline
  \end{tabular}
 \end{center}

The non-Sugawara dimensions $\Delta^{(0)}$ and spins $\ell^{(0)}$ are just $\Delta^{(0)} = h^{(0)} + \bar{h}^{(0)}, \ell^{(0)} = h^{(0)}- \bar{h}^{(0)}$. States are evenly divided between $\ell^{(0)} = + | \ell^{(0)}| $ and $\ell^{(0)} = - | \ell^{(0)}|$ at each weight, and the occupation numbers for the $(\mu, \bar{\mu}) = (0, \frac{1}{2})$ representations are the same as for $(\frac{1}{2}, 0)$.  To get the full characters one multiplies the non-Sugawara Virasoro characters $\chi_h(\tau)$ (i.e. generated by the modes of $T^{(0)} = T- T^{(\rm sug)}$) by the Weyl characters $\chi_\lambda^{(k)}(\tau,z)$, which in this case are\footnote{See eg \cite{DiFrancesco:1997nk}, eqs (14.176) and (15.244).}
\be
\chi_\lambda^{(1)}(\tau,z) = \frac{1}{\eta} \sum_{m \in \mathbb{Z}+\lambda} q^{m^2}  y^m .
\ee

After some trial and error, we find that the corresponding flavored partition function is reproduced by
\be
Z(\tau, \bar{\tau}, z, \bar{z}) = \frac{1}{4|\eta|^{6}} \sum_{a,b,a',b'=0}^1 (-1)^{a b' + a' b} \left|\theta\left[^a_b\right](\tau,\frac{z}{2})\right|^4 | \theta[^{a'}_{b'}](\tau,0)|^2 ,
\ee
in Jacobi/Erderlyi notation $\theta_1 = \theta[{1 \atop 1}], \theta_2 = \theta[{1 \atop 0}], \theta_3 = \theta[{0 \atop 0}], \theta_4 = \theta[ { 0 \atop 1}]$. 
As this candidate partition function is half integrally modded, it is a little unfamiliar. A natural guess is that it arises as a $\mathbb{Z}_{2}$ orbifold of a fully modular invariant theory.\footnote{There is actually a history of extremal theories arising in such a fashion \cite{Dixon:1988qd,Benjamin:2015ria,Harrison:2016hbq}.} Indeed it is possible to project this onto a fully modular invariant partition function. Taking the unflavored expression for simplicity,
\es{modinvproj}{
Z^{\rm(inv)}(\tau, \bar{\tau})&=\frac{1}{2}\left(Z(\tau, \bar{\tau}, 0,0)+Z(\tau+1, \bar{\tau}+1, 0,0)+Z(-1/(\tau+1), -1/(\bar{\tau}+1), 0,0)\right)
}
Unfortunately, we have not been able to identify a theory corresponding to \eqref{modinvproj}. It is straightforward to check by exhausting the possibilities that the central charge and the number of spin-1 conserved currents (11) is not consistent with this partition function being associated with a pure Sugawara theory for some Lie algebra.

While other choices for the quantization of spin are less conventional, they are still of some interest.\footnote{For instance \cite{ZamParafermion}.}  Another possibility we have considered is that the non-Sugawara part of the spin, i.e. $h^{(0)}-\bar{h}^{(0)}$, is an integer or a half-integer.  Because of the contribution to the weight from the Sugawara part of the stress tensor, in this case the states in the $(\frac{1}{2}, 0)$ and $(0,\frac{1}{2})$ representations have quarter-integer spins.  Performing the linear programming analysis making this Ansatz for the spins, we find not just a unique solution but in fact a family of solutions given by the following occupation numbers:
\begin{center}
\footnotesize
\begin{tabular}{|ccc|}
\hline
$ (\mu, \bar{\mu}) $ & $(\Delta^{(0)} , |\ell^{(0)} |)$ & $d$ \\
\hline 
$(0,0)$  & $(0,0)$ & 1 \\
\hline
 & $(1, 0)$ & 16 \\
 & $(1, 1 ) $ &12 \\
 \hline
 & $(2,0)$ & 36 \\
 & $(2, 1  ) $ & 32 \\
 & $(2, 2 ) $ & 20 \\
 \hline
 \end{tabular}
 \begin{tabular}{|ccc|}
 \hline
$ (\mu, \bar{\mu}) $ & $(\Delta^{(0)} , |\ell^{(0)} | )$ & $d$ \\
\hline 
 $(\frac{1}{2},0) $  & $(\frac{1}{2}, 0)$ & $4+x$  \\
 & $(\frac{1}{2}, \frac{1}{2})$ & $4-x$ \\
 \hline
 & $(\frac{3}{2}, \frac{1}{2})$ & $24+6x$ \\
 & $(\frac{3}{2}, 1)$ & $24-6x$ \\
 & $(\frac{3}{2}, \frac{3}{2} )$ & $4-x$  \\&&\phantom{s} \\
 \hline
  \end{tabular}
  \begin{tabular}{|ccc|}
  \hline
$ (\mu, \bar{\mu}) $ & $(\Delta^{(0)} , |\ell^{(0)} |)$ & $d$ \\
\hline 
 $ (\frac{1}{2} , \frac{1}{2}) $ & $(\frac{1}{2}, 0)$ & $8-2x$  \\
 & $(\frac{1}{2}, \frac{1}{2})$ & $2x$ \\
 \hline
 & $(\frac{3}{2}, \frac{1}{2})$ & $12 x$ \\
 & $(\frac{3}{2}, 1)$ & $48-12x$ \\
 & $(\frac{3}{2},\frac{3}{2}) $ & $2x$ \\
 &&\phantom{s} \\
 \hline
  \end{tabular}
 \end{center}

\begin{figure}[ht]
\includegraphics[width=\linewidth]{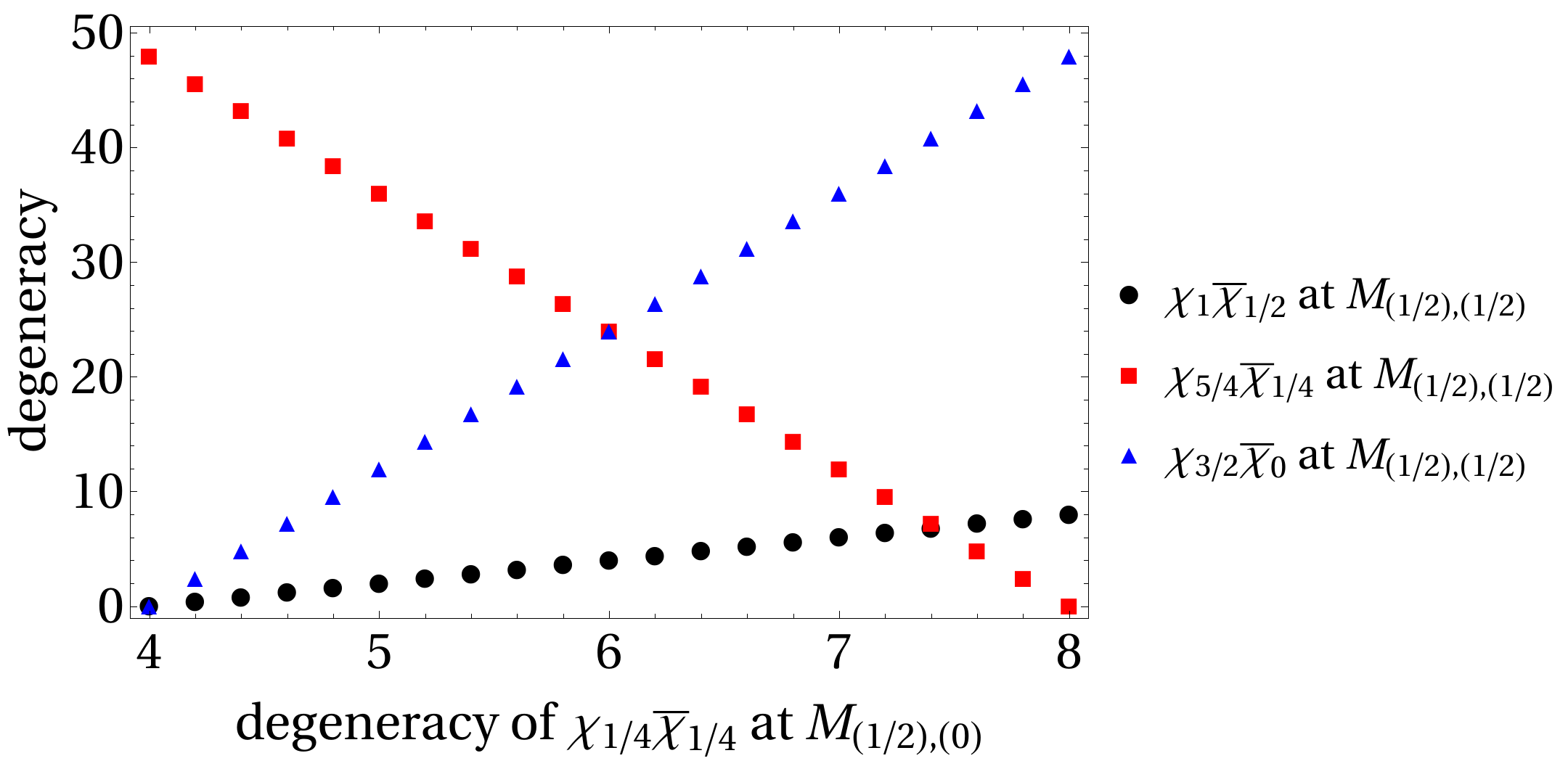}
\caption{The linear programming analysis finds a range of possible partition functions if we allow the physical spins to take quarter integer values.  When we fix one of the degeneracies $d$ by hand, in this case that of the weight $(\mu, \bar{\mu}, h^{(0)}, \bar{h}^{(0)})= (\frac{1}{2}, 0 , \frac{1}{4}, \frac{1}{4})$, all other degeneracies become uniquely determined, so that we find a one-parameter family of solutions.  The degeneracy on the x-axis here is $4+x$ in the notation used in the text.}
\label{fig:c3quarterspin}
\end{figure}

These partition functions satisfy crossing for all $x$. This one-parameter family is shown graphically in Fig. \ref{fig:c3quarterspin}, where we perform the linear programming analysis with the degeneracy $d$ of the $(\mu, \bar{\mu})= (\frac{1}{2} ,0), (h^{(0)}, \bar{h}^{(0)}) = (\frac{1}{4}, \frac{1}{4})$ chosen by hand and look at how several other degeneracies depend on this choice.  By inspection of the above table, demanding that all occupation numbers be non-negative integers imposes $x\in \{ 0,1,  \dots, 4\}$.  It would be interesting to know if all or any of these partition functions correspond to underlying physical CFTs.  

The different values of $x$ here correspond to partition functions that have the same spectrum of dimensions $\Delta= h + \bar{h}$, but which can be distinguished by their representation content, i.e. through the ``flavored'' partition function.\footnote{They are similar in this respect to multiple different CFTs at $c=24$ that have the same spectrum but different underlying symmetries \cite{Schellekens:1992db}.}

\subsection{Constraints on Representation Content}

In this final subsection, we will consider the question of what representations are forced to be present in a theory. The gravitational AdS dual of any such constraints would imply that even if certain representations were not present among the perturbative degrees of freedom in some theory, they would have to be present non-perturbatively.  The strongest condition one might try to prove is that all theories have all representations present.  This would however be too ambitious since there are simple counter-examples, but one might still try to prove restrictive constraints on which representations can be absent.  We will only be able to take a very modest step in this direction and prove some simple results for $SU(2)$.  For instance, without referring to numerical methods, we will prove that an $SU(2), k=1$ partition function either has all representations, or else its flavored partition function factorizes into a Sugawara theory partition function times a non-flavored partition function, assuming left-right symmetry.

We begin by proving this $k=1$ result. 
The flavored partition function splits into four
representations
\begin{equation}
  \label{eq:su2k1reps}
  M(\tau, \bar \tau) = \left(
  \begin{array}{cc}
    M_{(0),(0)}(\tau, \bar \tau)& M_{(\frac{1}{2}),(0)}(\tau, \bar \tau)\\
    M_{(\frac{1}{2}),(0)}(\tau, \bar \tau)& M_{(\frac{1}{2}),(\frac{1}{2})}(\tau, \bar \tau)
  \end{array}
  \right) .
\end{equation}
Modular invariance requires
\begin{equation}
  \label{eq:su2k1transform}
  M(-\frac{1}{\tau},-\frac{1}{\bar \tau}) = SM(\tau, \bar \tau)S~,
\end{equation}
with $S$ matrix
\begin{equation}
  \label{eq:su2k1Smat}
  S = \frac{1}{\sqrt{2}} \left(
  \begin{array}{cc}
    1 & 1 \\
    1 & -1
  \end{array}
\right)~.
\end{equation}
  Because there are only two (assuming $(\frac{1}{2},0)$ and $(0,\frac{1}{2})$ are symmetric) different nontrivial representations, and the modular transformation manifestly forces at least one to be present, we can delete only one of them.  
What if we set
$M_{(\frac{1}{2}),(0)}(\tau, \bar \tau)=M_{(0),(\frac{1}{2})}(\tau,
\bar \tau)=0$? In this case, the $(1,2)$ entry of matrix equation
(\ref{eq:su2k1transform}) is
\begin{equation}
  \label{eq:su2k1missing}
  \frac{1}{2}\left( M_{(0),(0)}(\tau, \bar \tau)
    -M_{(\frac{1}{2}),(\frac{1}{2})}(\tau, \bar \tau) \right) = 0 .
\end{equation}
The two diagonal representations have to be the same and therefore the ``non-Sugawara'' $\tau$-dependence of the flavored partition function is just an overall flavor-independent prefactor $M_{(0),(0)}(\tau, \bar{\tau})$ that factors out.

Similarly, if we set $M_{(\frac{1}{2}, \frac{1}{2})}(\tau,\bar{\tau})=0$,  then the $(2,2)$ entry of (\ref{eq:su2k1transform}) is $M_{(\frac{1}{2}),(0)}(\tau,\bar{\tau}) = \frac{1}{2} M_{(0),(0)}(\tau, \bar{\tau})$, so again the non-Sugawara $\tau$-dependence factors out completely.  In this case, the residual ``Sugawara'' matrix is just a symmetric holomorphic plus anti-holomorphic matrix, i.e. {\tiny $\left( \begin{array}{cc} 2 & 1 \\ 1 & 0 \end{array} \right) = \left( \begin{array}{cc} 1 & 1 \\ 0 & 0 \end{array} \right) +\left( \begin{array}{cc} 1 & 0 \\ 1 & 0 \end{array} \right)$. }

Beyond $SU(2)$ $k=1$, similar arguments can be used to somewhat narrow down the
possible combinations of representations in any partition function
with non-abelian currents.  Speficially, we can prove for SU(2) any $k$,
the partition function factorizes into a Sugawara partition function
times a flavor-independent partition function if only diagonal
representations are allowed.

We again begin with the transformation rule
\begin{align}
\tag{\ref{eq:su2k1transform}}
  M(-\frac{1}{\tau},-\frac{1}{\bar \tau}) = SM(\tau, \bar \tau)S~,
\end{align}
where now the transformation matrix is
\begin{align}
S_{(l) (l^\prime)} = \sqrt{\frac{2}{k+2}} \sin\left(\frac{\pi}{k+2} (l+1)(l^\prime+1)\right) .
\end{align}
Since we allow only diagonal
representations, we can write
\begin{align}
M_{(l) (r)}(\tau, \bar \tau) &= \delta_{(l) (r)} f_l , \\
M_{(l) (r)}(-\frac{1}{\tau},-\frac{1}{\bar \tau}) &= \delta_{(l) (r)} g_l~,
\end{align}
for some arbitrary functions $f_l$ and $g_l$.
The matrix equation can be written as
\begin{align}
g_\alpha S_{\alpha \beta}=S_{\alpha \beta} f_\beta .
\end{align}
For $\beta = 0$,
\begin{align}
g_\alpha \sin\left(\frac{\pi}{k+2} (\alpha+1)\right) =f_0 \sin\left(\frac{\pi}{k+2} (\alpha+1)\right)
\end{align}
so $g_\alpha = f_0$ for all $\alpha$ unless
$\sin\left(\frac{\pi}{k+2} (\alpha+1)\right) = 0$. But the unitary bound
$\alpha<k+1$ does not allow this to happen. So all $g_\alpha$ should
be equal.  Therefore, all diagonal representations
$M_{(l) (l)}(\tau, \bar \tau)$ have to be equal, and the $\tau$-dependence $f_0(\tau)$ of $M(\tau, \bar{\tau})$ completely factors out.

\section{Discussion and Future Directions}

One of the main goals of this paper has been to demonstrate how systematic numeric bootstrap techniques can be applied to flavored partition functions.  We have considered several specific analyses, but there are many more that could be done. Here we will discuss a few potential future directions.

Some of the analyses we have discussed raise questions that could be answered with improved numeric efficiency so that the results could converge to the optimal bound.  One such case is the bound on the charge-to-mass ratio, where improved accuracy at large $c$ could more firmly establish the large $c$ scaling of the bound.  Another case is the application of our nonabelian extremal methods to larger $k$ and larger symmetry groups.  As either of these gets larger, the convergence rate becomes slower and so we have focused on the most efficient case, $SU(2)$ at level $k=1$, to demonstrate that here the extremal functional method can be used to determine the full partition function of the theory maximizing the gap in the neutral sector.  It is interesting that the point maximizing the gap has integer occupation numbers, and it would be interesting to know if this is part of a general pattern or just an exceptional case.  Our preliminary analysis of $SU(2)$ at level $k=2$ has not converged well enough to answer this question, but perhaps this would be possible with additional innovations  or more computing power.  Of course, if it turns out that integer occupation numbers is a generic feature of maximal gap spectra, it would be interesting to understand the underlying reason.  As part of this question, one might consider whether the gap should be maximized in just the neutral sector or in several charged sectors.  

Having integer occupation numbers is a necessary but not sufficient condition for a partition function to have an underlying CFT.  Generally, it would be interesting to develop more techniques for determining a CFT once its partition function is known.  One way is simply to use the regular bootstrap but restricting all dimensions to those that appear in the partition function.  Usually, this is a significant improvement since it reduces the regular bootstrap problem to a linear programming problem; however, for rational theories, the large degeneracy at each level severely mitigates how helpful this additional information is.  Another possible approach one could try would be to use the partition function formulated as the four-point function of twist operators, $Z \propto \< \sigma_2 \sigma_2 \sigma_2 \sigma_2\>$, to include the partition function together with $\< \phi \phi \phi \phi\>$ and the ``mixed'' correlator $\< \phi \phi \sigma_2 \sigma_2\>$ for some local operator $\phi$. 

One could also try to make contact at $c=24$ with the Schellekens classification \cite{Schellekens:1992db} in terms of Neimeier lattices, by rederiving this constraint using only the flavored modular bootstrap. The modular bootstrap alone cannot constrain the number of currents, since they simply contribute a constant to the partition function, but a constant no longer satisfies the correct transformation law after flavoring.  

Looking farther afield, one of the main motivations for developing a proof of the transformation law (\ref{eq:txn}) in terms of background fields was that this might be easier to generalize.  There are many theories in 2d with higher spin currents, and one could generalize our derivation to such cases.  The correlators of higher spin currents do not have a simple universal generating functional like spin-one currents do, but their correlators are severely constrained by holomorphicity and crossing, and recursion relations are known in many cases. Potentially, one could work out the transformation rule in a case-by-case fashion.  More ambitiously, one could try to generalize to $d>2$.  The very interesting recent work \cite{Shaghoulian:2016gol} on a sort of modular invariance for lens spaces in higher dimension is tantalizing from this point of view.  Again, one would face the issue that correlators of currents in $d>2$ are not universal, but one could nevertheless try to obtain a constraint on the partition function in terms of the data in the $\< J(x_1) \dots J(x_n)\>$  correlators.

Somewhat more abstractly, one of the appealing features of understanding the flavored partition function better is that, by turning on background fields, we are exploring constraints beyond the class of those that can be seen by inserting local operators.  There are many such constraints on CFTs that are invisible in the standard bootstrap; the partition function itself can be though of as one such generalization since mapping to the torus (equivalently, inserting twist operators $\sigma_2$) involves imposing new boundary conditions, and adding background fields is another kind of generalization.  It would be very interesting to understand what additional constraints could be obtained by imposing crossing symmetry of correlators in the presence of background fields.  Understanding the transformation law (\ref{eq:txn}) as a statement about crossing symmetry for the four-point function $\< \sigma_2 \sigma_2 \sigma_2 \sigma_2\>$ would be a useful warm-up case and could potentially give insight into how to think about more general correlators.

\section*{Acknowledgments}

We thank  Scott Collier, Shamit Kachru, Jared Kaplan, Emanuel Katz, and Per Kraus for useful conversations.  ED was supported  in part by the National Science
Foundation under grant NSF-PHY-1316699 and by the Stanford Institute for Theoretical Physics and in part by the Simons Collaboration Grant 488655 on the Non-Perturbative Bootstrap.  ALF and YX were supported in part by the US Department of Energy Office of Science under Award Number DE-SC-0010025, and ALF was supported in part by  the  Simons  Collaboration  Grant  on  the  Non-Perturbative Bootstrap.

\begin{appendices}

\end{appendices}

\section{Path Integral Modular Transformation}
\label{app:PItxn}

In this appendix, we review how diffeomorphism invariance and rigid rescalings imply the relation
\be
Z_{\rm PI}\left( \frac{a \tau+b}{c\tau+d}, \frac{c z}{c \tau+d} \right) = Z_{\rm PI}(\tau,z).
\ee
We begin with invariance of the path integral measure:
\be
d \Psi e^{ - S_{\tau_a, \tau_b}[\Psi]} = d\Psi' e^{ - S_{\tau'_a, \tau'_b} [\Psi']}.
\label{eq:meas2}
\ee
Here, $\Psi$ are all the fields of the CFT, and to keep track of the torus before and after conformal transformations we have introduced $\tau_a, \tau_b$ for two of its corners (i.e. the four corners are at $0,\tau_a, \tau_b$, and $\tau_a+\tau_b$). Under rescalings, the operators $\CO$ and parameters $\tau_a, \tau_b$ transform as
\be
\CO(w,\bar{w}) \rightarrow \CO'(w,\bar{w}) = \lambda^{-h} \bar{\lambda}^{-\bar{h}}\CO(\lambda^{-1} w,  \bar{\lambda}^{-1} \bar{w}), \qquad (\tau_a, \tau_b) \rightarrow (\tau_a',\tau_b')= (\lambda \tau_a, \lambda \tau_b).
\ee
In particular, for a conserved current $J^\mu$, we have
\be
\int_{\tau_a, \tau_b} dw d\bar{w} J^w (w) = \int_{\tau_a, \tau_b} dw d\bar{w}\Big( \bar{\lambda}  J'^w(\lambda w)\Big) = \lambda^{-1} \int_{\tau'_a, \tau'_b} dw d\bar{w} J'^w(w) .
\ee
and consequently
\be
d \Psi e^{ - S_{\tau_a, \tau_b} [\Psi]  - \frac{i}{2\pi } \int_{\tau_a, \tau_b} dw d\bar{w} A_w J^w} = d \Psi' e^{- S_{\tau'_a, \tau'_b}[\Psi'] - \frac{i}{2\pi}  \int_{\tau'_a, \tau'_b} dw d\bar{w} \lambda^{-1} A_w J^w }.
\ee
Integrating both sides obtains the relation $(A_w, \tau_a, \tau_b) \cong (\lambda^{-1} A_w, \lambda \tau_a, \lambda \tau_b)$.

To obtain the transformation under $U: \tau\rightarrow \frac{\tau}{\tau+1}$, we take
\be
(\tau_a, \tau_b) = (\tau, \tau+1) \cong (\tau,1),
\ee
where the congruence $\cong$ follows from a large diffeomorphism cutting the torus along the line from 1 to $\tau+1$ and sewing it back to the line from $\tau+1$ to $\tau+2$.  By inspection of the chemical potential term $\frac{1}{2\pi i} \int_{\tau,1} dw d\bar{w} A_w J^w = 2 \pi i {\rm Im}(\tau) A_w \bar{J}_0$, we read off that
\be
A_w = - i \frac{\bar{z}}{2 {\rm Im}(\tau)}.
\ee
Finally, we take $\lambda = (\tau+1)^{-1}$, so $(\tau_a',\tau_b')=(\frac{\tau}{\tau+1}, 1)$ and $A_w' = (\tau+1) A_w$.  Therefore,
\be
\bar{z}' = 2 i {\rm Im}(\tau') A_w' = \frac{\bar{z}}{\bar{\tau}+1}.
\ee
The transformation under $T: \tau\rightarrow \tau+1$ is trivial, since $\tau$ and $\tau+1$ are related by a large diffeomorphism without any need for a rescaling, so $\lambda=1$, and neither $A_w$ nor $z$  transform.  All other modular transformations are generated from $T$ and $U$.

\section{A ``Systematic'' Treatment to Multivariate Problems} 
\label{sec:multivariate}

The bootstrap of flavored partition function introduces another
continuous quantum numbers $Q$ in addition to the scaling dimension of
$\Delta$. Unlike the unflavored bootstrap where the problem is
rigorously converted to a semidefinite programming problem, bootstrap
problems with more than one variables do not have a simple and
rigorous conversion to semidefinite programming problems. One can
choose to discretize the second variable $Q$ and hope that the bound
converges at very small $\delta Q$. However, the bound obtained in
this way is not rigorous. The linear functional can be negative in
between discrete $Q$'s or at large enough $Q$. 

\subsection{Multivariate Positive Definite Functionals}

Whether any real positive semidefinite polynomials (PSD) can be
written as sum of squares of real polynomials (SOS) is known as the
Hilbert's 17th problem. Hilbert himself proves the special case for
univariate polynomials is true. But for multivariate polynomials it is
later proven that PSD is a sum of squares of real rational
functions. We do not like rational functions because we have much less
numerical control over them than polynomials.

Although we cannot find a clean SOS representation of multivariate
PSD, if we only consider the subset of strictly positive polynomials
we can still represent them by SOS in the following cases: 

{\bf Workaround 1: multiply by a common denominator} $p(x_1,x_2)$ is
{\bf positive definite} polynomial (PD, also denote as
$p(x_1,x_2) > 0$) then $p_g(x_1,x_2) = (1+x_1^2+x_2^2)^g p(x_1,x_2)$
is a sum of square of polynomial (SOS) for some $g$. \cite{reznick1995uniform}

{\bf Workaround 2: region is bounded } For a compact region
$\mathcal S$ defined by $f_i(\vec x)\geq 0$ over set of function $f_i$, any
polynomial strictly positive in $\mathcal S$ can be written as the
following form
\begin{align}
  p = \sum_I s_I(\vec x) f_{i_1} f_{i_2}\ldots
\end{align}
where $s_I(\vec x)$ are sum of squares. $I$ denotes some
combinations of $f_i$'s. \cite{powers2000polynomials}

The hope is that PD can approximate PSD well enough so that in
practise we can still resort to SOS. Numerically, solvers like SDPB
never give nonnegative polynomials with exact zeros, so in practise we
never actually encounter any counterexamples. Another reason to
be hopeful is from the proof that PSD can be approximated as closely
as desired by SOS \cite{lasserre2007sum}.

There is a possible loophole -- the positive region of the polynomial
has to be bounded. In unflavored case the region that is frequently
used is $\Delta>\Delta_{\star}$, which is a rare special case of
unbounded region. In practise, there is risk of not covering the full
space of PD. Although not rigorously, one can hope that by multiplying
the $(1+x_1^2+x_2^2)^g$ factors of higher and higher $g$ we lose less
and less.

\subsection{Multivariate Problems and SDPB}
In this subsection we discuss how to rewrite the semidefinite
polynomial programming with 2 variables into a form suitable for SDPB
\cite{Simmons-Duffin:2015qma} solver. 
SDPB solves univariate ``Polynomial Matrix Program'' (PMP) question
stated as follows: 
\begin{align}
  &\text{maximize } y_0 + \sum_n b_n y_n \nn \\
  &\text{such that } M_j^0(x) + \sum_n y_n M_j^n(x) \geq 0 \nn \\
  &\text{for all } x\geq 0 \text{ and } 1\leq j \leq J.
\end{align}
where $M$ matrices are symmetric matrices of polynomials of $x$.

In SDPB, the PMP question is internally mapped to an SDP question
since $ M_j^0(x) + \sum_n y_n M_j^n(x) \geq 0$ if and only if
\begin{align}\label{eq:oneVariableFunctionalMatrixForm}
  M_j^0(x) + \sum_n y_n M_j^n(x) =~ \tr \left[ Y_{A} Q_{A}(x) \right] +
     x\, \tr \left[ Y_{B}  Q_{B}(x)\right]
\end{align}
for some $Y_A, Y_B \geq 0$. This equation is also the (2.8) of 1502.02033.

We are instead trying to solve the problem for two variable
cases.
Here for modular bootstrap we are in the special case where the
symmetric matrices $M_j$ are one by one, in other words, single
polynomials $p_j$. For simplicity here we only deal with this one
dimensional case. Generalization to more dimensions and more variables
is very easy.
The question is stated as follows:
\begin{align}
  &\text{maximize } y_0 + \sum_n b_n y_n \nn \\
  &\text{such that } p_j^0(x_1,x_2) + \sum_n y_n p_j^n(x_1,x_2) \geq 0 \nn \\
  &\text{for all } x_1\geq 0 \text{ and all $x_2$ and } 1\leq j \leq J.
\end{align}
Since SDPB only allows one variable to be bounded we cannot add more
constraints on the variables. The $x_1>0$ is needed in SDPB because we
usually choose the input $\Delta \geq \Delta_*$. The second
variable can be the U(1) charge $Q$, which is not constrained to be
positive number. If one does want to bound the second variable one can
make change of variable. 
We use the symbol $F_j$ to represent the linear functional
\begin{align}\label{eq:twoVarLinearFunctional}
  F_j(x_1,x_2) \equiv p_j^0(x_1,x_2) + \sum_n p_j^n(x_1,x_2) y_n
\end{align}
Similar to the univariate case, we assume that $F_j\geq 0$ is
equivalent to finding $Y_{A,j},~Y_{B,j} \geq 0$, so that
\begin{align}
  F_j(x_1,x_2)  =~ \tr \left[ Y_{A,j} Q_{A}(x_1,x_2) \right] +
     x_1 \tr \left[ Y_{B,j}  Q_{B}(x_1,x_2)\right] 
\end{align}
Here we introduce ``bilinear basis'' $\vec{q}(X)$ so that $Q(X) = \vec{q}
\vec{q}^T$ spans the space of polynomials of $X$.
An easy example of bilinear basis is $\vec q(x) = \{ 1,x,x^2,\ldots\}$.
The bilinear basis of two or more variables can be factored out as a
kronecker product of bilinear bases of each single variables
\begin{align}
  Q_A (x_1,x_2) &= Q_{A1}(x_1) \otimes Q_2(x_2) \nn\\
  Q_B (x_1,x_2) &= Q_{B1}(x_1) \otimes Q_2(x_2)
\end{align}
We define $d_i$ to be the $x_i$ degree of the polynomial $F$. the
dimensions of the matrices $Q$ are
\begin{align}
  &{\rm dim} Q_{A1} = \delta_{A1} = [d_1/2]+1 \nn \\
  &{\rm dim} Q_{B1} = \delta_{B1} = [(d_1-1)/2]+1 \nn \\
  &{\rm dim} Q_{2} = \delta_{2} = [d_2/2]+1
\end{align}
After factoring out $Q_A$ and $Q_B$ the function $F_j$ is written as
\begin{align}\label{eq:twoVariableFunctionalMatrixForm}
  F_j(x_1,x_2) =~ \tr \left[ Y_{A,j}
  \big( Q_{A1}(x_1) \otimes Q_2(x_2)\big) \right] +
  x_1 \tr \left[ Y_{B,j} \big( Q_{B1}(x_1) \otimes
  Q_2(x_2)\big)  \right] 
\end{align}
Since a polynomial is fixed if we know its value at $(d+1)$ different
points, we can simply evaluate the above equation at $(d_2+1)$ values
of $x_2$ in order to reduce the equation to have only one variable
$x_1$\footnote{We have not investigated what choices of the $x_{2,k}$s are optimal.  In practice, we have taken them to be $x_{2,k} = 2^{k-1}$. }
\begin{align}
  &F_{j,k}(x_1) = F_j(x_1,x_{2,k}) =~ \tr \left[ Y_{A,j}
  \big( Q_{A1}(x_1) \otimes Q_2(x_{2,k})\big) \right] +
  x_1 \tr \left[ Y_{B,j} \big( Q_{B1}(x_1) \otimes
  Q_2(x_{2,k})\big)  \right]
  \label{eq:Fjkdef}
\end{align}
The above $(d_2+1)$ equations are equivalent to
(\ref{eq:twoVariableFunctionalMatrixForm}).
In the following we omit the $j$ index because the same equation works for
all $j$.
Now the form is already in single variable and is very close to the
form of (\ref{eq:oneVariableFunctionalMatrixForm}). The only difference is the numerical matrices
$Q_2(x_{2,k})$. Here we can play a trick by shuffling the $(d_2+1)$
equations with linear combination 
\begin{align}
  \label{eq:orthorgonalizeSingleVariableBasis}
  &\sum_l \alpha_{kl} F_l =  \tr\left[ Y_{A,j}  \big(
    Q_{A1}(x_1) \otimes \sum_l \alpha_{kl} Q_2(x_{2,l})
      \big) \right] + (\text{B part})
\end{align}
for some dimension $(d_2+1)$ square matrix $\alpha_{kl}$.
In fact, the space of symmetric $Q_2$ matrices is only $(d_2+1
= 2\delta_2 -1)$ dimensional space since it spans the space $(d_2+1)$
dimensional polynomials. That means we can always find some
$\alpha_{kl}$ which picks up the the orthornormal
basis of the polynomial space.
Further we can perform an arbitrary $GL_{\delta_2}$ transformation on
$Q_2$
\begin{align}
  Y &\mapsto G YG^{-1} \nn \\
  \sum_l \alpha_{kl} Q_2(x_{2,l}) &\mapsto G^{-1} \sum_l \alpha_{kl}
                                    Q_2(x_{2,l}) G
\end{align}
so that the orthornormal basis maps to the symmetric matrix basis
\begin{align}\label{eq:conjectureCombinationQ}
  G^{-1} \sum_l \alpha_{kl} Q_2(x_{2,l}) G = E^{r(k) s(k)}
\end{align}
where $E^{rs} = \delta_i^r \delta_j^s + \delta_j^r \delta_i^s$.
Then
\begin{align}\label{eq:matrixElementForm}
  \sum_l \alpha_{kl} F_l = \tr \left[ Y_A Q_{1A}(x_1) \otimes E^{r(k)
  s(k)}\right] +  x_1\tr \left[ Y_B Q_{1B}(x_1) \otimes E^{r(k)
  s(k)}\right]
\end{align}
Compared to (\ref{eq:oneVariableFunctionalMatrixForm}), we can turn
double variable programming of polynomial into single variable
programming of symmetric polynomial matrices by substitution
\begin{align}
  \label{eq:matrixBasisDefinition}
  M_j^0 &= \sum_k \sum_l E^{r(k) s(k)} \alpha_{kl}
          P_j^0(x_1,x_{2,l})\nn \\
  M_j^n &= \sum_k \sum_l E^{r(k) s(k)} \alpha_{kl}
          P_j^n(x_1,x_{2,l})
\end{align}
Since $Q_2$ span a $(d_2+1 = 2\delta_2 -1)$ dimension space it means
only the diagonal and next-to-diagonal elements will be nonzero. If
further we only have even powers of $x_2$, the matrices will be
diagonal.

The procedure defined from
(\ref{eq:orthorgonalizeSingleVariableBasis}) to
(\ref{eq:matrixBasisDefinition}) is not the most efficient algorithm
to obtain a single-variable matrix basis to input in
(\ref{eq:oneVariableFunctionalMatrixForm}).  In practice, the algorithm we actually follow is computationally more straightforward, and is as follows.  We can rewrite (\ref{eq:oneVariableFunctionalMatrixForm}) as
\begin{align}
  M_j(x_1)_{tu} =
  Y_{A,j}^{rs,tu}\, Q_{A1}(x_1)_{rs} + (\text{B parts}) ~,
\end{align}
where $r,s,t,u$ are matrix indices.  Given $M_j(x_1)$, SDPB can optimize this single-variable problem.  Rather than obtaining $M_j(x_1)$ by the procedure outlined above, we can instead start directly from equation (\ref{eq:Fjkdef}), which says
\begin{align}
  \label{eq:equationToDecideOneVariableMatrix}
  F_{j,k}(x_1) = Y_{A,j}^{rs,tu}\, Q_{A1}(x_1)_{rs}\,
  Q_2(x_{2,k})_{tu} + (\text{B parts}) ~.
\end{align}
Combining  the two equations
\begin{align}
  F_{j,k}(x_1) = M_j(x_1)^{tu}\, Q_2(x_{2,k})_{tu} ~.
\end{align}
The point is that both $F_{j,k}(x_1)$ and $Q_2(x_{2,k})_{tu} $ are known, so  $M$ can be obtained by solving the above equation, after which it can be fed into SDBP.
Concretely, first flatten the $(t,u)$ indices
$ \alpha := (t_\alpha,u_\alpha)$
to put $M_{tu}$ into
$(\delta_2+1)(\delta_2+2)/2$ dimensional array ${\bf M}_\alpha$ and $Q_{k,tu}$
into $(d_2+1)\times(\delta_2+1)(\delta_2+2)/2$ array
$Q_{k,\alpha}$. Then solve the linear equations for ${\bf M}_\alpha$. 

To be concrete, in this paper we explicitly choose the flatten map
\begin{align}\label{eq:flattenMap}
  {\bf M} := \big(\text{diag of $M$},~\text{next-to-diag},~\cdots \big)~.
\end{align}
A specific solution can be found by taking the SVD decomposition of $Q$,
\begin{align}
  \label{eq:svdQ}
  Q = U \, W \, V^T
\end{align}
where diagonal matrix
$$W = \left( \ \widetilde W \quad O \  \right)$$ has the same rank $(d_2+1)$ as $Q$.
Then take
\begin{align}
\label{eq:svdQ}
  B^{T} = V \, W^\prime \, U^T
\end{align}
where
$$W^\prime = \left( 
  \begin{array}{c}
    \widetilde W^{-1} \\ O
  \end{array} \right)
$$
is the pseudo inverse of $W$. Contract $B^T$ both sides of
(\ref{eq:equationToDecideOneVariableMatrix}),
% the left hand side is
\begin{align}
  B^T {\bf F} &= V \, W^\prime \, U^T Q {\bf M}
  = V \, W^\prime \, U^T U \, W \, V^T {\bf M}
  = V \, \left(
  \begin{array}{cc}
    I_{d_2+1}& O \\ O & O
  \end{array}
  \right) \, V^T {\bf M}
\end{align}
where it's useful to block-decompose the unitary matrix $V$ as

$$%\begin{align}
  V = \left(
  \begin{array}{cc}
    V_{11}&V_{12}\\ V_{21}&V_{22}
  \end{array}
  \right)~.
$$%\end{align}
Continue to simplify the right hand side,
\begin{align}
  V^TB^T{\bf F} &= \left(
  \begin{array}{cc}
    V_{11}&V_{12}\\ O & O
  \end{array}
                        \right) {\bf M} \nn\\
  \big( (V_{11}^T)^{-1} \  O  \big)V^TB^T{\bf F} &= \left( \  I_{d_2+1} \quad
                                 (V_{11}^T)^{-1}V_{21}^T \  \right){\bf
                                 M} ~.
\end{align}
From (\ref{eq:matrixBasisDefinition}) we know that there are only
$(d_2+1)$ independent elements are unique. Any dependent element can
be removed by a $GL_{\delta_2}$ transformation defined by
\ref{eq:conjectureCombinationQ}. We choose ${\bf M}$ such that only
the first $(d_2+1)$ elements are zero, and the solution can be written
as
\begin{align}
  {\bf M} = \big( (V_{11}^T)^{-1} \  O  \big)V^TB^T{\bf F}~.
\end{align}
Since $F_j(x_1,x_2)$ is the linear functional of input
polynomials defined by (\ref{eq:twoVarLinearFunctional}). Our single
variable input matrices should be substituted in the same way, leading to
\begin{align}
  {\bf M}_{j}^n &= \big( (V_{11}^T)^{-1} \  O  \big)V^TB^T{\bf
                  P}_{j}^n \nn \\
   {\bf M} &\equiv {\bf M}_{j}^0 \, y_0 + \sum_n {\bf M}_{j}^n \, y_n
             \nn \\
  \left( {\bf P}_{j}^n \right)_k &\equiv {\rm Flatten}\left(  p_j^n(x_1,x_{2,k}) \right)
\end{align}

Finally, we inverse (\ref{eq:flattenMap}) to put flattened array
${\bf M}_j^n$'s back into matrix $M_j^n$'s.

\section{$k=2, SU(2)$ Analysis}
\label{app:su2k2}

\begin{figure}[t!]
\begin{center}
\includegraphics[width=0.6\textwidth]{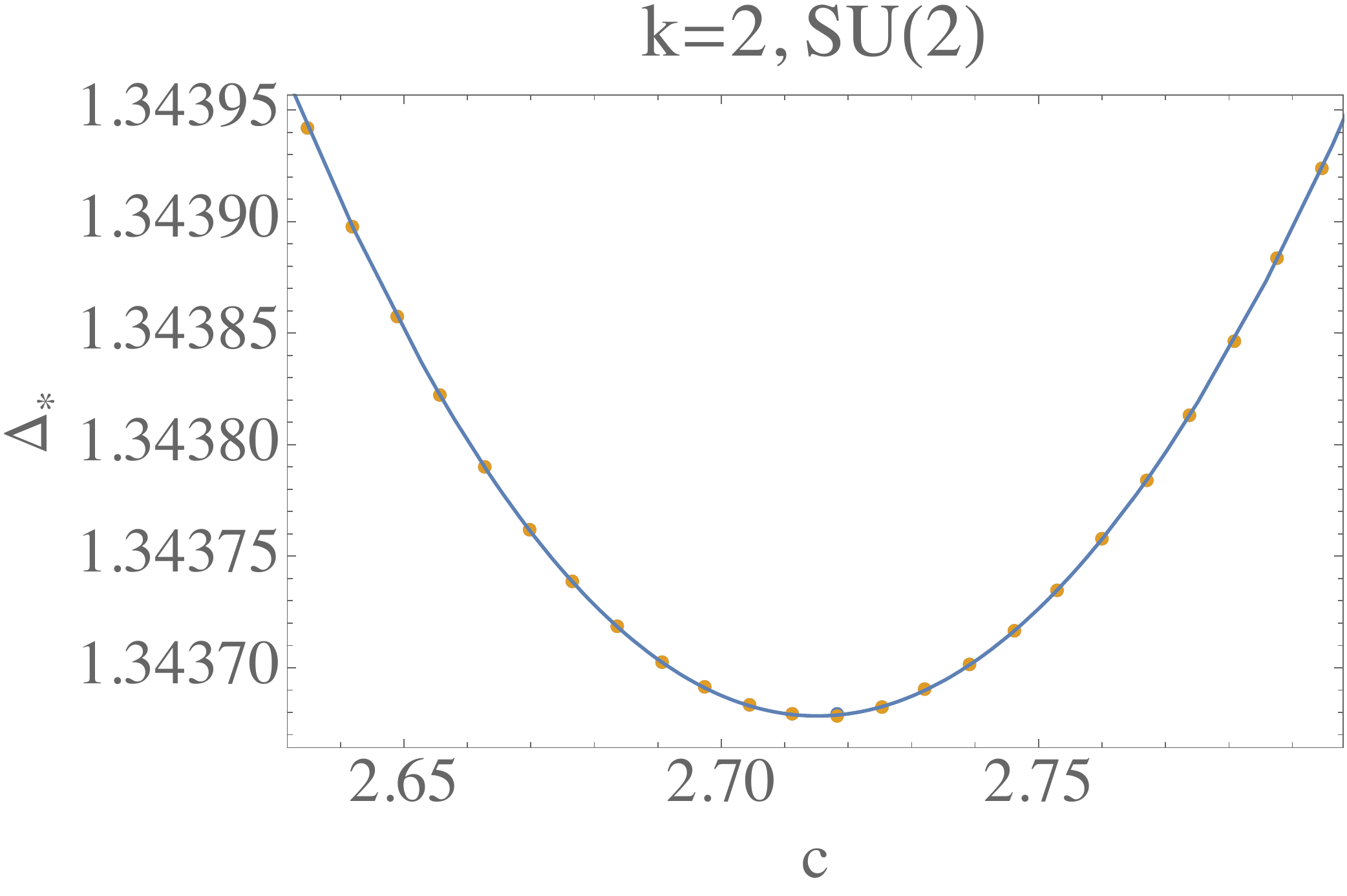}
\caption{Bound on the gap $\Delta_*$ to the lightest neutral state for $SU(2)$ at level $k=2$.  The bound is minimized at $c \approx 2.715$.}
\label{fig:su2k2-bound}
\end{center}
\end{figure}

Here we present some preliminary results on our methods applied to the group $SU(2)$ at level $k=2$.  Our results are qualitatively similar to the $k=1$ case, though with  worse numeric accuracy due to the slower convergence.  In Fig. \ref{fig:su2k2-bound}, we show the bound on the gap $\Delta_*$ to the lightest neutral state in the theory, which is minimized to be $\Delta_* \approx 1.344$ at $c \approx 2.715$.  

\begin{figure}[t!]
\begin{center}
\includegraphics[width=0.99\textwidth]{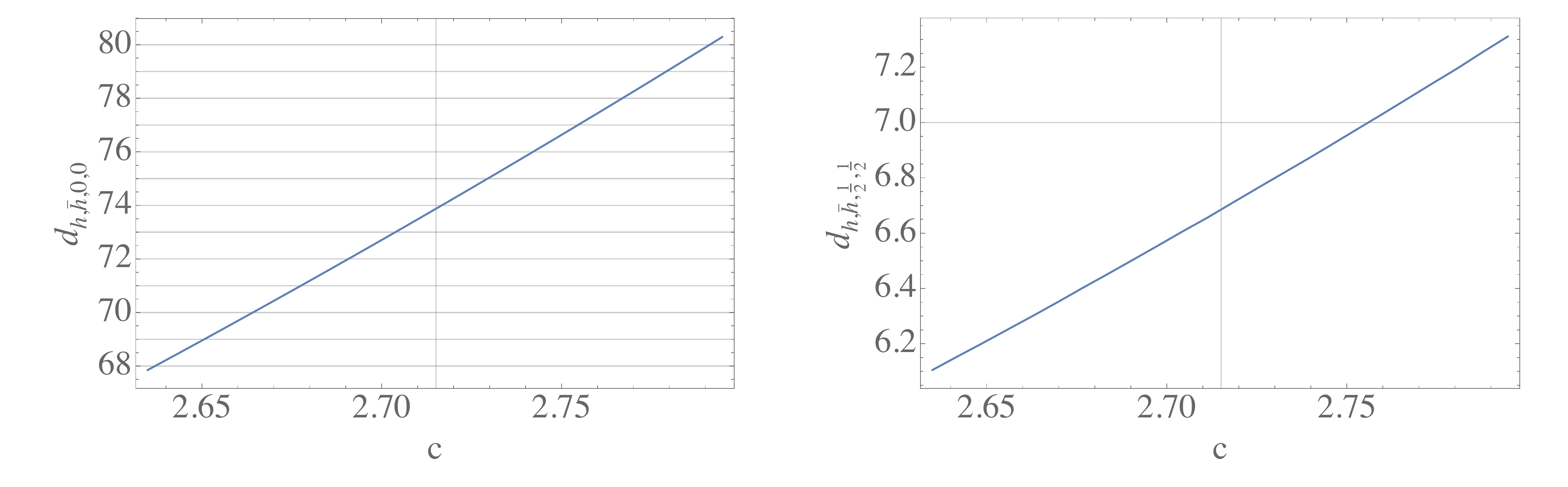}
\caption{Occupation numbers for the lightest neutral (left) and charged (right) states from the extremal functional analysis with $SU(2)$ at $k=2$ as a function of $c$.  The optimal bound is at $c\approx 2.715$, indicated by a vertical line; horizontal lines are shown at integers.  }
\label{fig:su2k2-ds}
\end{center}
\end{figure}

Unfortunately, at the point where the bound is minimized, the occupation numbers from our analysis for some of the lowest few states are not particularly close to integers.  It is not clear whether this indicates that such a point is not associated with an underlying CFT or if we simply have not converged to sufficient precision.  The occupation numbers for the lightest neutral state and charged state are shown as a function of $c$ in Fig. \ref{fig:su2k2-ds}.  The lightest neutral state is close to $d=74$, however the lightest charged state, which is even lighter  is relatively far from the nearest integer, $d \approx 7$.  Another possibility is that one ought to maximize the gap in not only the neutral sector but also in one or more charged sectors; it would be interesting to pursue this or other conditions further.

\newpage
\phantom{s}

\newpage

\bibliographystyle{utphys}
\bibliography{ModFlav}

\end{document}